\documentclass{SCIS2026}

\usepackage{hologo}
\usepackage{pifont}
\usepackage{makecell}
\usepackage{multirow}
\usepackage{multicol}
\usepackage{tablefootnote}
\usepackage{xcolor}
\usepackage{url}
\usepackage{wasysym}
\usepackage{threeparttable}
\usepackage{marvosym}
\usepackage{wrapfig}
\usepackage{xspace}
\usepackage{ulem}
\usepackage{graphicx}
\usepackage{capt-of}
\definecolor{gainGreen}{RGB}{34,139,34}
\definecolor{lossRed}{RGB}{190,45,45}
\definecolor{tieGray}{RGB}{105,105,105}

\DeclareRobustCommand{\JY}[1]{\textcolor{black}{#1}}
\DeclareRobustCommand{\JYY}[1]{\textcolor{black}{#1}}

\newcommand{\SysName}{\textsc{MazeRunner}\xspace}

\AtBeginDocument{
  }
    
\begin{document}
\ArticleType{RESEARCH PAPER}
\Year{2026}
\Month{}
\Vol{}
\No{}
\DOI{}
\ArtNo{000000}
\ReceiveDate{}
\ReviseDate{}
\AcceptDate{}
\OnlineDate{}
\AuthorMark{}
\AuthorCitation{}

\title{MazeRunner: Nonlinear Task and Clue Orchestration for LLM-driven Black-Box Automated Penetration Testing}
{MazeRunner: Nonlinear Task and Clue Orchestration}
\author[1]{Zhenyuan Li}{{lizhenyuan@zju.edu.cn}}
\author[1]{Yi Jiang}{}
\author[1]{Junjie Cheng}{}
\author[1]{Yaokun Li}{}
\author[2]{Jing Qiu}{}
\author[1]{Shouling Ji}{}

\address[1]{Zhejiang University, Hangzhou 310058, China}
\address[2]{Guangzhou University, Guangzhou 510006, China}

\abstract{
Penetration testing constitutes a resource-intensive yet indispensable component of modern network security. Although large language models (LLMs) have demonstrated substantial potential in automating security auditing, existing LLM-driven agents remain largely confined to executing end-to-end workflows within simplified, linear scenarios. Real-world black-box engagements are fundamentally different: the underlying attack graph is initially unknown and must be incrementally constructed from environmental feedback. Each observation may expose multiple plausible attack branches, failures are often semantically ambiguous, and decisive clues may be separated by vast action horizons. Consequently, current autonomous agents repeatedly fall into depth-first traps, misattribute failures, and forget previously discovered evidence.
We present \SysName, an autonomous penetration testing system based on a three-agent task and clue orchestration framework. \SysName separates global task orchestration, context-intensive execution, and failure-oriented review, while maintaining task states and environmental evidence in a persistent Task \& Clue Cache. This design enables the system to revise failed actions, recover missing prerequisites, switch between attack branches, and correlate clues across distant execution stages.
We evaluate MazeRunner on 10 recently released Hack The Box targets under a common budget of 20 million LLM tokens per system–target run and a protocol designed to prevent target-specific solution leakage. With Claude Sonnet 4.5, MazeRunner completes 47.7\% of the annotated subtasks, compared with 36.2\% for PentestGPT-V2 and 34.2\% for Claude Code. MazeRunner obtains user-level or higher access on 6 targets, including root access on 2, whereas each same-model baseline obtains user-level access on 2 targets and fails to reach root access. Analysis of the complete execution traces further shows that MazeRunner explores a broader range of attack branches and achieves higher shell-acquisition efficiency per token than the same-model baselines.
}


\maketitle

\section{Introduction}
As a proactive and rigorous security assessment methodology, penetration testing (pentesting) plays a pivotal role in both industrial and academic domains. According to Fortune Business Insights, the global annual expenditure on pentesting reached \$2.74 billion in 2025~\cite{fortune2025pentest}. Nevertheless, traditional pentesting relies heavily on human expertise, suffering from prohibitive labor costs, low efficiency, and poor scalability, thereby fundamentally limiting its widespread deployment~\cite{shen2025pentestagent}. Therefore, automated pentesting and threat assessment solutions have garnered extensive attention from both researchers and practitioners.

Although often conceptualized as a sequence of standard phases, black-box pentesting initiates with only a target and an ultimate objective, lacking a known execution path. The underlying attack graph is initially opaque and must be incrementally synthesized as new environmental clues, such as unknown services or credentials, are discovered. This renders the exploitation process inherently dynamic, requiring frequent task generation, lateral pivoting, and strategic recalibration based on emerging evidence~\cite{happe_can_2025}. Moreover, environmental feedback is highly ambiguous: execution failures do not necessarily invalidate a chosen path, while apparent successes or promising clues can easily trigger deceptive ``rabbit holes.'' Because critical evidence required for late-stage decisions may be acquired dozens of steps earlier~\cite{chen2025breaking}, relying on an ever-growing LLM context window is fundamentally inadequate. Instead, effective autonomous pentesting mandates the integration of a persistent and verifiable external state architecture.


\begin{table*}[tbp]
\centering
\caption{Comparison of \SysName with existing LLM-based automated pentesting systems}
\label{tab:systemComparison}
\footnotesize
\resizebox{\textwidth}{!}{
\begin{tabular}{l|cc|cc|cc}
\toprule
{System} & {\makecell{Auto-\\matic}} & {\makecell{Scenario Scope}} & {\makecell{Independent\\Task/Clue}} & {Failure/Fault Handling} & {\makecell{Multi-Agent\\Architecture}} & {\makecell{Tool/Target\\Interaction}} \\
\midrule
AutoPT~\cite{wu2025autopt} & $\times$ & Vuln. Exploiting & Vuln. only & Vuln. Switching & Single & CLI \\
PentestAgent~\cite{shen2025pentestagent} & $\times$ & Vuln. Exploiting & $\times$ & $\times$ & ``Recon-Plan-Execute'' & CLI + Manual \\
CAI~\cite{mayoral2025cai} & $\times$ & Open & $\times$ & $\times$ & Specialist Agents & CLI + Manual \\
PentestGPT~\cite{deng2024pentestgpt} & $\times$ & Open & Task only & Manual & Single & Manual \\
TermiAgent~\cite{mai2025shell} &\checkmark & Open & Task only & $\times$ & Single & CLI \\
Incalmo~\cite{singer2025incalmo} & \Circle & Multi-host & Task + Clue & $\times$ & ``Plan-Execute'' & CLI + Scripts \\
PentestMCP~\cite{zhai2025pentestmcp} &\checkmark & Vuln. Exploiting & Task only &  \makecell{Plan Switching} & ``Recon-Enum-Exploit'' & MCP + CLI \\
Trinity~\cite{para2026trinity} & \checkmark & Multi-host & Clue only & Self-Healing Retry & \makecell{``Plan-Guard-\\Execute-Observer''} & CLI \\
Cochise~\cite{happe_can_2025} & \checkmark & AD-specific & Task only & Branch Switching & ``Plan-Execute'' & CLI \\
PentestGPT-V2~\cite{deng2026makes} & \checkmark & Open & Task only & Review + Restart & Dynamic Sub-agents & CLI \\
\midrule
\SysName & \checkmark & Open & Task + Clue & Review + Branch Selection & ``Plan-Execute-Review'' & CLI \\
\bottomrule
\end{tabular}
}
\end{table*}

\begin{wrapfigure}{r}{0.49\columnwidth}
  \centering
  \vspace{-\intextsep}
  \includegraphics[width=\linewidth]{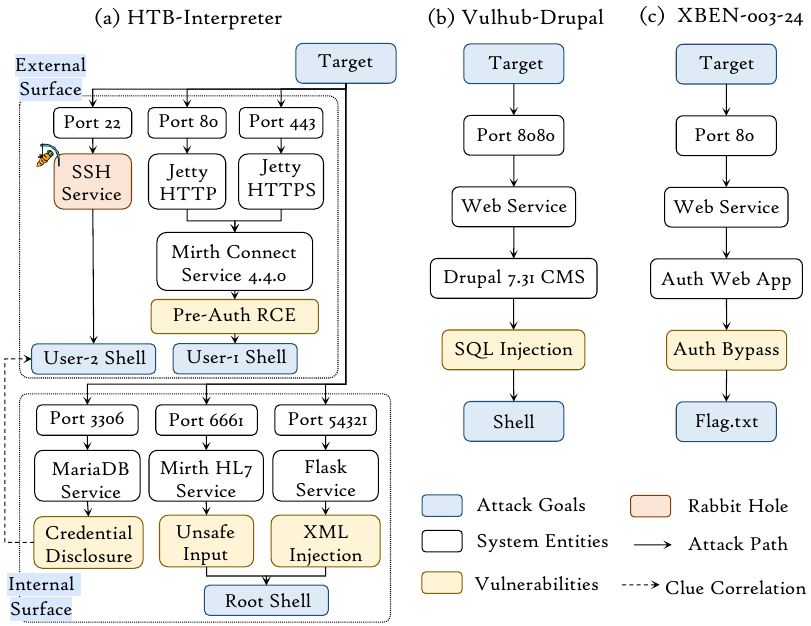}
  \caption{Complexity comparison of diverse pentesting targets
  }
  \label{fig:motivatingExample-1}
\end{wrapfigure}

\begin{figure}[!t]
    \centering
    \includegraphics[width=\linewidth]{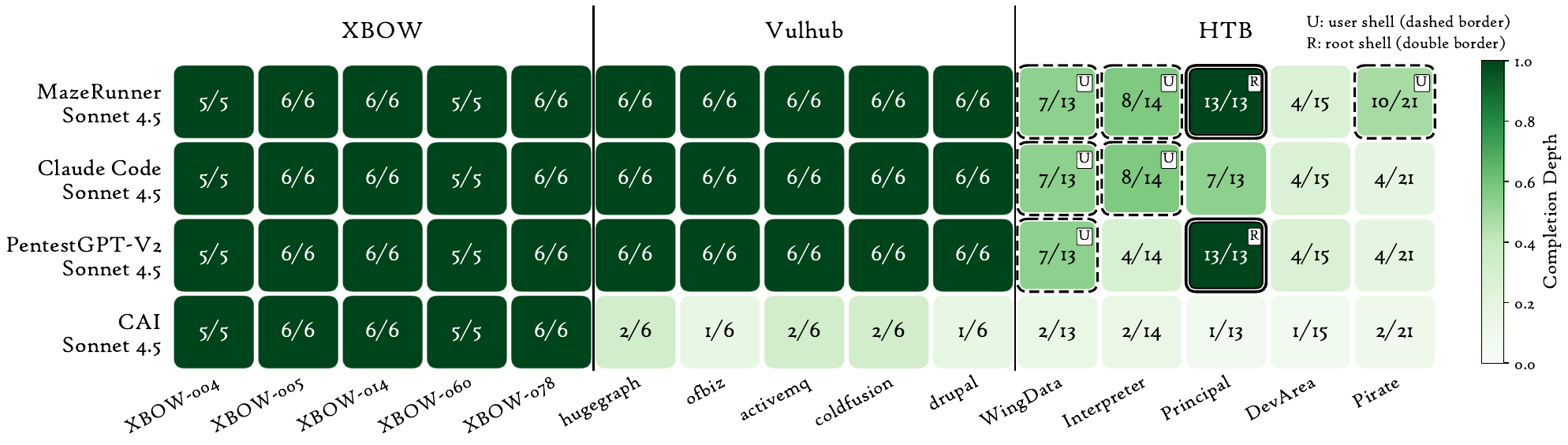}

    \begingroup
    \setlength{\abovecaptionskip}{3pt}
    \setlength{\belowcaptionskip}{-6pt}
    \caption{Preliminary performance comparison across various targets}
    \label{fig:pre-study1}
    \endgroup
\end{figure}

Early automation efforts in pentesting mainly focused on vulnerability scanning and attack surface analysis~\cite{metasploit, jing2024revisiting, searchsploit}. Due to the absence of robust analytical reasoning and dynamic decision-making capabilities required in unconstrained scenarios, the automation of subsequent, more complex exploitation phases has remained largely stagnant. The advent of Large Language Models (LLMs) has catalyzed new opportunities, inspiring recent works to leverage LLMs for automating these advanced stages~\cite{singer2025incalmo, deng2024pentestgpt, shen2025pentestagent, mayoral2025cai}.
As summarized in Table~\ref{tab:systemComparison}, these pioneering frameworks have successfully demonstrated the feasibility of utilizing LLMs to interpret raw security tool outputs, map recognized vulnerabilities to corresponding exploits, and autonomously generate actionable payloads. Operationally, the majority of these systems interface with target environments by executing standard security tools through Command Line Interfaces (CLI). By leveraging the vast security knowledge embedded within their foundational models and employing basic heuristic planning, these systems excel at executing atomic operations. Consequently, they have achieved notable success in automating relatively straightforward, linear attack paths, particularly within standardized or isolated Capture-The-Flag (CTF) environments, as illustrated in Figure~\ref{fig:motivatingExample-1} and Figure~\ref{fig:pre-study1}.

\JY{These unique domain characteristics manifest in existing LLM-driven
pentesting frameworks as three recurring operational bottlenecks:}
1) Deficient Orchestration \& The ``Depth-First Trap'': Absent rigorous supervision, the stochastic nature of LLMs leads to unpredictable task orchestration. When a task fails, agents frequently hyper-focus on that local branch. As execution context accumulates, this disproportionate anchoring drives repeated, unproductive attempts, inevitably culminating in goal drift, hallucinated tasks, and extra overhead.
2) Contextual Amnesia: The finite context window of LLM agents cannot accommodate the massive volume of feedback generated during iterative environmental interactions, causing tasks and clues to be ignored or forgotten in long-term engagements.
3) Error Misattribution: LLMs struggle with accurate state tracking and root-cause analysis. Following a failure, their decision-making degrades; they often erroneously link the failure to unrelated tasks, leading to the premature abandonment of viable attack trajectories.
As evidenced by our study (\S\ref{sec:background} and \S\ref{sec:evaluation}), these bottlenecks constrain both the depth and breadth of LLM-driven automated pentesting.

To address these limitations, we propose {\SysName}, a multi-agent automated pentesting framework architected specifically for task and clue orchestration within complex, black-box targets. At its core, the framework formalizes tasks and clues as fundamental primitives managed within a persistent, independent cache. This ensures strict state consistency to mitigate contextual amnesia and task redundancy. {\SysName} operates through a continuous, collaborative ``Strategist-Executor-Reviewer'' agent cycle. Specifically, we introduce a dedicated Reviewer Agent tasked with diagnosing execution failures to provide actionable, corrective feedback; this mechanism allows the system to escape the ``Depth-First Trap'' and enhances the precision of subsequent task selection/correction. Simultaneously, we expand the operational scope of the Strategist Agent beyond localized planning, empowering it to orchestrate global penetration strategies and dynamically recalibrate the task graph in response to execution anomalies. The Executor Agent is responsible for operationalizing these strategies into concrete environmental interactions while harvesting critical runtime clues. Ultimately, this closed-loop system synthesizes LLM-driven reasoning with deterministic procedural logic to enable resilient, adaptive, and zero-intervention pentesting.

To evaluate \SysName, we conducted experiments across 10 Hack The Box (HTB) targets released between February and April 2026\footnote{HTB, https://www.hackthebox.com/}. The targets cover Linux and Windows systems and span easy, medium, and hard difficulty levels. To enable a controlled comparison, we impose a common maximum budget of 20 million LLM tokens on each system–target run and prohibit the use of target-specific online write-ups. 
With Claude Sonnet 4.5, MazeRunner achieves a subtask completion rate of 47.7\%, compared with 36.2\% for PentestGPT-V2 and 34.2\% for Claude Code. It achieves strictly greater subtask progress than each same-model baseline on 6 of the 10 targets. MazeRunner obtains user-level or higher access on 6 targets and reaches root access on 2 of them, whereas each baseline obtains user-level access on 2 targets and reaches root access on none. Analysis of the execution traces indicates that the baselines frequently expend their budgets on repeated local attempts or terminate before exploring viable alternative branches. By maintaining persistent task and clue states and introducing an independent failure-review stage, MazeRunner reduces such execution loops and improves shell-acquisition efficiency. In one representative run, it obtains root access using 79 commands, of which 7 are classified as invalid. The main contributions of this work are as follows:

\begin{itemize}
  \setlength{\itemsep}{0pt}
  \setlength{\parskip}{0pt}
  \setlength{\parsep}{0pt}
  \item We propose \SysName, an autonomous pentesting system built upon a multi-agent task and clue orchestration framework. By separating strategic task orchestration, context-intensive execution, and failure-oriented review, MazeRunner supports adaptive navigation of nonlinear attack paths in complex black-box environments.

    \item We design a persistent Task \& Clue Cache that maintains task dependencies, execution states, environmental evidence, and cross-stage clue associations. Combined with an independent Reviewer Agent, this design supports failure diagnosis, prerequisite recovery, branch switching, and long-horizon evidence reuse.

    \item We evaluate MazeRunner on 10 recently released HTB targets under a common resource budget. With Claude Sonnet 4.5, MazeRunner achieves a 47.7\% subtask completion rate and obtains user-level or higher access on 6 targets, including root access on 2. The corresponding baselines reach 36.2\% and 34.2\% completion and obtain user-level access on 2 targets without reaching root. MazeRunner also achieves broader attack-path exploration and higher shell-acquisition efficiency per token in the evaluated runs.
\end{itemize}

\section{Background and motivation}
\label{sec:background}

\subsection{Automated pentesting}

penetration testing (pentesting) is an authorized simulated attack designed to evaluate the security of systems, networks, or applications. Traditionally, this process has relied heavily on human experts. However, this manual paradigm is severely constrained by prohibitive costs, a chronic shortage of skilled professionals~\cite{isc22025workforce}, and systemic risks introduced by human oversight within complex workflows~\cite{uptime2025outage}. Furthermore, periodic manual testing is increasingly inadequate for securing rapidly evolving attack surfaces, driving the critical need for automated pentesting to enable continuous security validation.
Despite this urgent demand, automating pentesting remains a formidable challenge~\cite{simon2024sok, skandylas2025automated}. In practice, pentesting necessitates invoking diverse tools through an iterative exploration cycle of ``task orchestration, execution, information analysis, and knowledge/tool retrieval'' under highly uncertain black-box conditions. Consequently, effective automation demands not merely the mechanical execution of attacks, but rather sophisticated, dynamic attack planning~\cite{wang2025unified}.

Table~\ref{tab:systemComparison} summarizes representative LLM-powered pentesting systems, all of which aim to enhance efficiency and automation. However, the majority of these frameworks fall short of achieving end-to-end automation. For instance, tools such as CAI~\cite{mayoral2025cai} and PentestGPT~\cite{deng2024pentestgpt} function merely as conversational assistants, providing heuristic advisory recommendations rather than executing actions. Conversely, systems like PentestAgent~\cite{shen2025pentestagent} and AutoPT~\cite{wu2025autopt} focus primarily on vulnerability exploitation, lacking the comprehensive capability to handle other essential phases of the pentesting lifecycle. \JY{TermiAgent~\cite{mai2025shell}improves memory management and exploit execution, but provides limited support for global task and clue organization.}
Furthermore, frameworks such as Cochise~\cite{happe_can_2025} and Incalmo~\cite{singer2025incalmo} achieve automation only within constrained scenarios. Specifically, Incalmo addresses task planning in multi-host environments but remains partially dependent on pre-defined scripts, whereas Cochise is exclusively tailored for Active Directory targets. \JY{PentestMCP~\cite{zhai2025pentestmcp} primarily emphasizes standardized tool integration, while Trinity~\cite{para2026trinity} introduces stateful multi-agent reasoning for network pentesting.} Notably, general-purpose coding agents (e.g., Claude Code) have recently demonstrated a considerable degree of automated penetration capability. Building upon these paradigms, PentestGPT-V2~\cite{deng2026makes} has largely realized a fully autonomous, zero-intervention pentesting workflow.

\subsection{Preliminary Study on Pentest Targets and Baseline}

\begin{wrapfigure}{r}{0.50\columnwidth}
    \centering
    \vspace{-\intextsep}
    \includegraphics[
        width=\linewidth
    ]{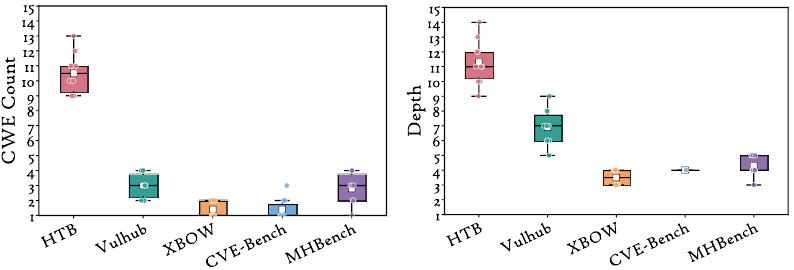}
    \caption{Complexity of Pentesting Benchmarks}
    \label{fig:Complexity-of-Benchmarks}
\end{wrapfigure}

Recent studies~\cite{wu2025autopt, shen2025pentestagent, deng2024pentestgpt, deng2026makes} demonstrate the potential of LLM agents in automated pentesting, particularly within simplified, ``CTF-like'' environments. As illustrated in Figure~\ref{fig:motivatingExample-1} (b, c), these tasks typically feature well-defined objectives and strictly linear progression paths, often reducing to a predictable sequence: a single open port, a specific service, an isolated vulnerability, a direct exploit, and the final target. Due to this structural simplicity, LLMs can maintain task focus without being diverted by environmental noise.
Figure~\ref{fig:Complexity-of-Benchmarks} further quantifies the complexity of various benchmarks, including HTB, Vulhub\footnote{Vulhub, https://vulhub.org/}, XBOW\footnote{XBOW, https://xbow.com/}, CVE-Bench\footnote{CVE-Bench, https://github.com/uiuc-kang-lab/cve-bench}, and MHBench\footnote{MHBench, https://arxiv.org/abs/2501.16466}, by comparing the average number of CWEs per target and the measured task depth. 
Task depth quantifies the cumulative number of requisite sub-tasks (e.g., port scanning, information gathering, and vulnerability exploitation) necessary to successfully compromise the target (a detailed breakdown is provided in Appendix Table~\ref{tab:gt-inter}).
The results indicate that HTB targets exhibit significantly higher complexity in both breadth and depth compared to other datasets. Consequently, when confronted with the non-linear trajectories of complex targets such as ``HTB-Interpreter'' (Figure~\ref{fig:motivatingExample-1} (a)), which present multiple potential paths and misleading rabbit holes, existing LLMs struggle.

\begin{figure*}[tbp]
  \centering
  \includegraphics[width=0.96\linewidth]{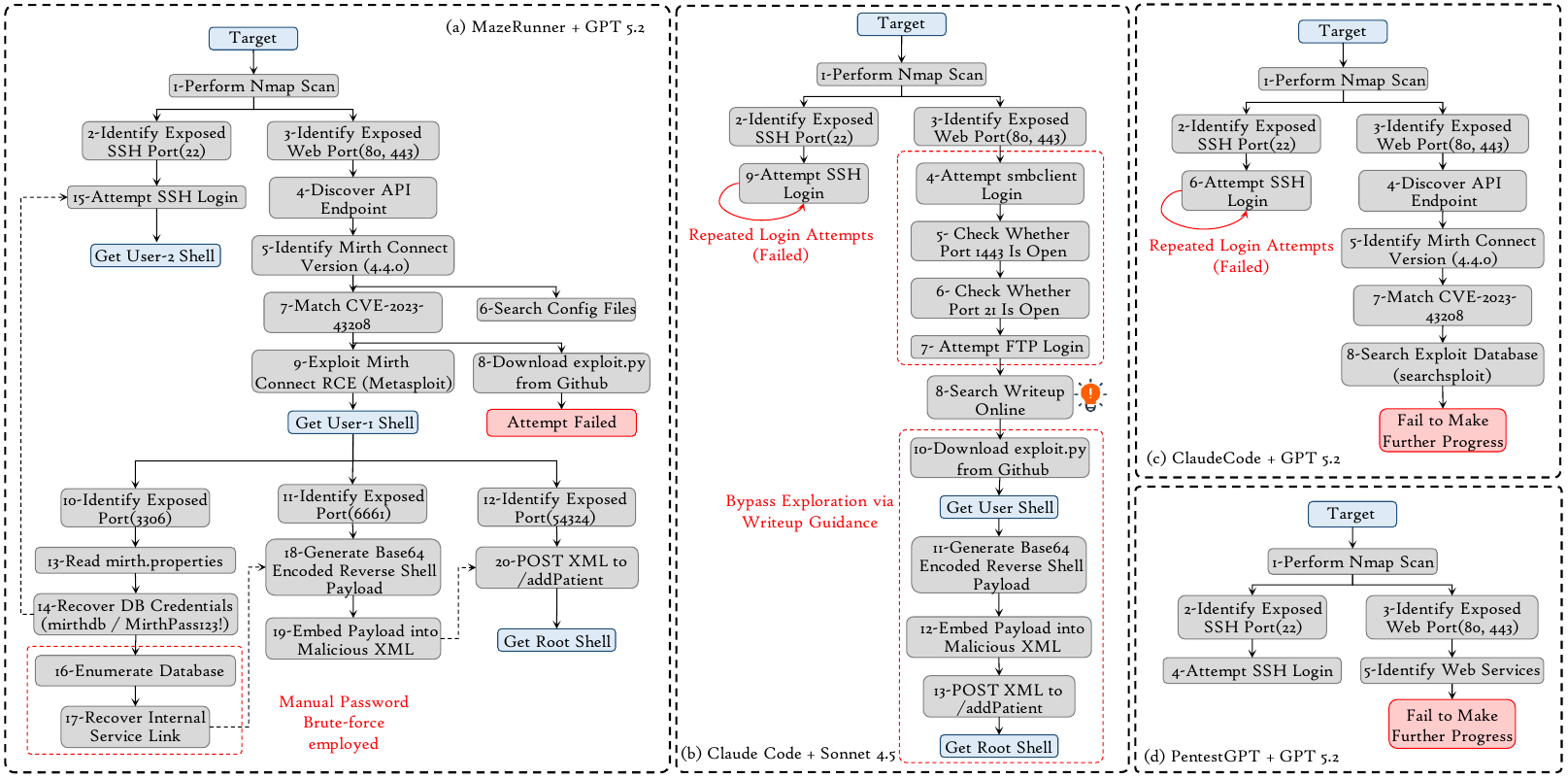}
  \caption{Illustrative diagnostic trajectories on HTB-Interpreter. These runs are excluded from the formal evaluation in Evaluation. Panel (a) additionally includes a human-assisted password-cracking step.}
  \label{fig:motivatingExample-2}
\end{figure*}

Figure~\ref{fig:pre-study1} summarizes the penetration efficacy of three representative baseline systems alongside our proposed \SysName across \JY{three} distinct environments: HTB (non-linear), Vulhub (linear), and XBOW(linear). For the underlying foundational model, we selected Claude Sonnet 4.5, a recent iteration widely adopted and highly performant across the aforementioned works.
As the results indicate, with the exception of CAI, which pauses for human instruction following initial reconnaissance, all evaluated systems, including Claude Code, successfully achieved full compromise on the Vulhub and XBOW targets. This high success rate is attributable to two factors: the relative structural simplicity of Vulhub and XBOW tasks, \JY{and the fact that many benchmark targets and their corresponding solutions significantly predate the model’s training data cutoff. During execution, we observed that the LLMs readily recognized these environments and rapidly retrieved known solutions from their parametric memory. However, this inherent advantage introduces severe evaluation bias and fails to generalize to zero-day or non-linear scenarios.}

Conversely, when evaluated against more complex and novel HTB challenges (specifically, new targets released in 2026), the performance of the baseline systems declined precipitously, yielding two shells on two targets. In contrast, \SysName enhanced overall penetration efficacy, successfully securing 1 root shell and 3 user shells. Comprehensive experimental details and further metrics are deferred to the Evaluation section.
It is crucial to note that the PentestGPT-V2 results reported in our study diverge significantly from the performance metrics presented in their original paper. This discrepancy arises because their evaluation methodology only accounted for the publication dates of official write-ups; however, their underlying Claude Code agent actively leverages internet search capabilities to retrieve third-party write-ups and exploit scripts, resulting in severe data contamination. To ensure a rigorous and fair evaluation, our experimental design explicitly instructed the baseline agents via system prompts to strictly avoid referencing external write-ups, and we meticulously verified the integrity of the resulting execution traces.

\subsection{Motivating Example}
To further illustrate the limitations of current systems when navigating complex, non-linear targets, we present a detailed case study of the ``HTB-Interpreter'' machine. As depicted in the comparative example in Figure~\ref{fig:motivatingExample-2}, we contrast the progression of existing baseline agents against our proposed \SysName framework within this identical complex pentesting environment.

\begin{wrapfigure}{r}{0.47\columnwidth}
  \centering
  \vspace{-\intextsep}
  \includegraphics[width=\linewidth]{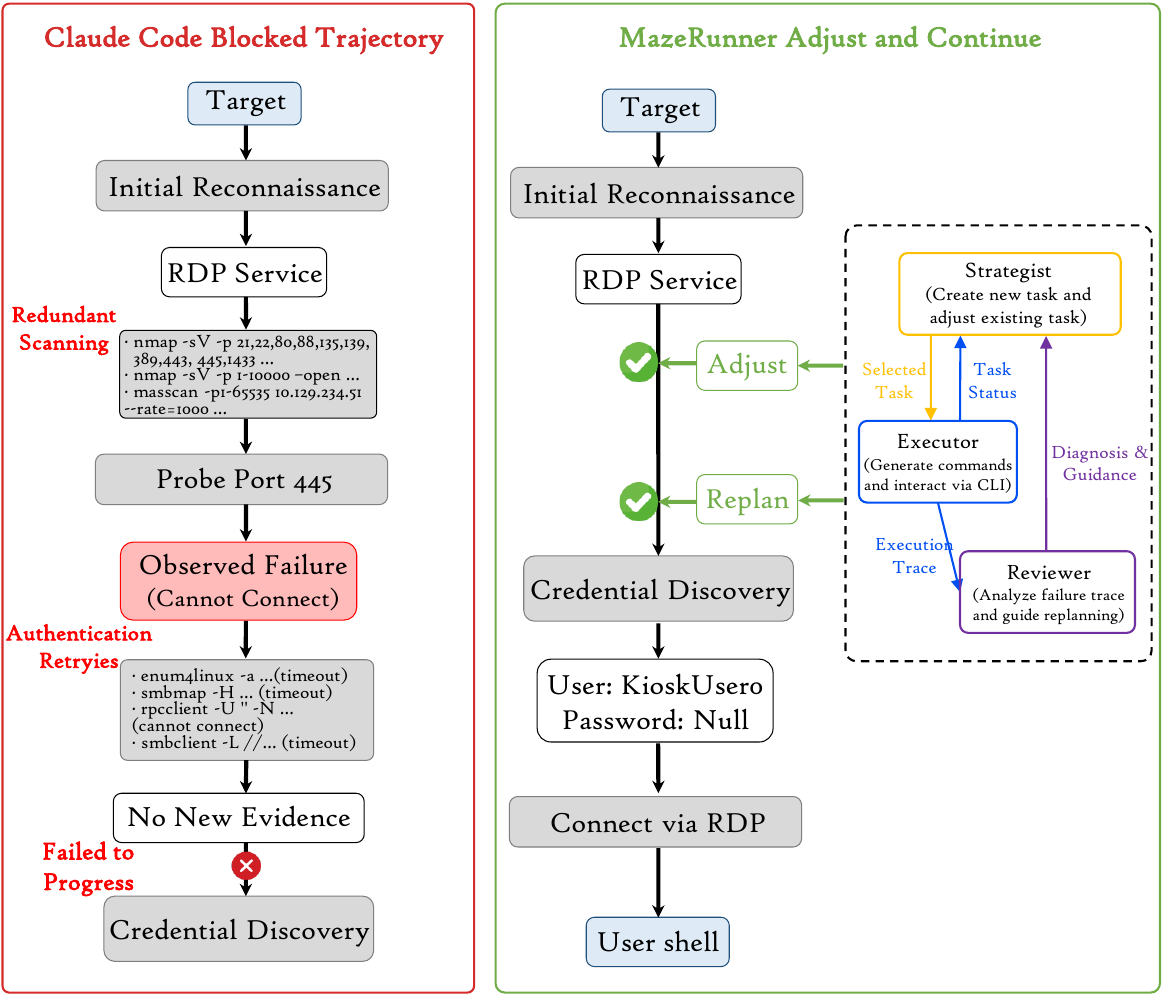}
  \caption{Redundant, hallucinated, inconsistent commands}
  \label{fig:case study}
\end{wrapfigure}

\JY{Overall, current automated pentesting systems exhibit significant limitations in task orchestration, state tracking, and failure recovery when handling long-horizon, black-box targets.} As shown in Figure~\ref{fig:motivatingExample-2} (b), (c), and (d), agents relying on standard LLM toolchains (e.g., Claude Code or PentestGPT-V2) are easily trapped in ``rabbit holes'' during exploration. This results in a multitude of redundant or even hallucinated tasks, as detailed in Figure~\ref{fig:case study} (``HTB-VulnEscape''). For instance, these agents expend critical resources on repeated, unsuccessful SSH login attempts or engage in irrelevant probing that diverges from the critical attack path (e.g., blindly testing unassociated FTP or SMB ports). Furthermore, agents such as Claude Code (Figure~\ref{fig:motivatingExample-2} (b)) frequently resort to retrieving online write-ups to bypass the genuine exploration phase. In scenarios devoid of readily available solutions, or upon encountering tool execution failures (Figure~\ref{fig:motivatingExample-2} (c) and (d)), these baseline systems halt completely and fail to make further progress. This exposes a severe deficiency in error recovery and adaptability within \JY{complex, non-linear benchmark} environments.

In contrast, as depicted in Figure~\ref{fig:motivatingExample-2} (a), \SysName demonstrates exceptional capability in dynamic decision-making, error-aware task backtracking, and cross-stage clue correlation. Following an initial failed exploitation attempt (Step 8: downloading a faulty exploit script), the system avoids falling into an infinite retry loop. Instead, it successfully backtracks and pivots to a viable alternative strategy (Step 9: deploying Metasploit). Crucially, after acquiring the initial user shell, the system exhibits robust long-term state maintenance. It systematically reads configuration files, recovers database credentials, and enumerates internal services. 
It is worth noting that steps 16 and 17 require executing a password brute-force attack. Since relying on the LLM to automate this process incurs prohibitive token overhead and temporal latency, we employed manual intervention for this specific task. This allowed us to bypass the mechanical bottleneck and continue evaluating the framework's overall end-to-end efficacy. Following this manual phase, the autonomous agent seamlessly resumes control, synthesizing the accumulated, fragmented cross-stage clues to construct the precise malicious XML payload required to ultimately secure Root access.

This contrast highlights that relying solely on rudimentary vulnerability matching and naive tool execution is profoundly insufficient for dynamic, \JY{non-linear} penetration scenarios. Endowing autonomous agents with the dual capacity to recognize errors and backtrack within uncertain environments, coupled with the ability to persistently maintain and integrate fragmented clues across multiple attack stages, is imperative for overcoming current bottlenecks in automated pentesting. Addressing this fundamental gap serves as the core motivation driving this research.

\section{\SysName design}

\label{sec:workflow}


\subsection{System Overview}
\label{sec:architecture}

Automated black-box pentesting is inherently challenging because the target's internal topology and the underlying attack graph are initially opaque. Furthermore, dynamically emerging clues, competing attack branches, and semantically ambiguous execution failures profoundly complicate persistent state tracking and strategic replanning. Despite recent advancements, current LLM-driven pentesting frameworks fundamentally struggle to navigate these non-linear environments, where complex branching frequently precipitates erratic exploration. Consequently, existing solutions remain structurally constrained by their reliance on continuous human intervention or rigid, predefined execution pipelines, severely impeding their autonomy and real-world generalizability.

\begin{figure*}[!t]
  \centering
  \includegraphics[width=0.84\linewidth]{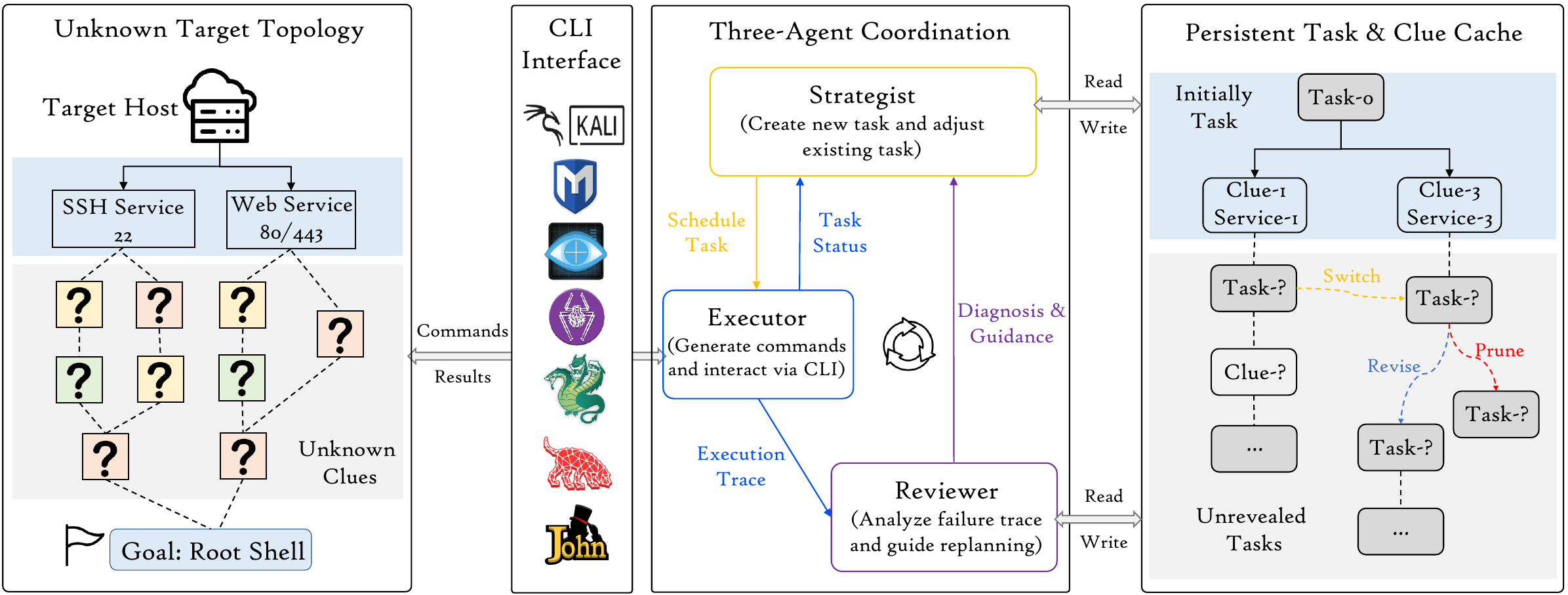}
  \caption{\SysName's architecture}
  \label{fig:architecture}
\end{figure*}

\SysName strives for full autonomy in black-box pentesting across complex, multi-stage scenarios. It deliberately eschews reliance on static strategies or predefined scripts~\cite{deng2026makes} that inherently constrain generalizability and practical applicability. Instead, the framework leverages LLMs for adaptive orchestration, decomposing high-level penetration objectives into dynamic tasks, such as code generation for environmental discovery and targeted intelligence retrieval. By designating tasks and clues as core primitives and managing them through an independent cache, \JY{\SysName maintains a verifiable investigation state while mitigating task redundancy, contextual forgetting, and hallucinations. The cache comprises a Task Graph for task dependencies and execution states, and a Clue Graph for structured environmental evidence.} Finally, the architecture employs three collaborative agents, as illustrated in Figure~\ref{fig:architecture}, to implement a ``strategist-executor-reviewer'' cycle, effectively balancing open-ended discovery with proactive error correction, thereby preventing the squandering of resources on unproductive trajectories.

\JY{Specifically, the Strategist decomposes the user-provided objective and maintains the evolving task structure. The Executor receives the selected task, translates it into CLI operations, and records execution evidence and newly discovered clues in the shared cache. When an execution is unsuccessful or inconclusive, the Reviewer analyzes the resulting trace and provides structured guidance to the Strategist. The Strategist then revises the current plan, adds missing prerequisite tasks, or switches to an alternative branch. Deterministic state transitions, such as dependency resolution and downstream blocking, are handled programmatically. Together, these components form a closed-loop workflow covering task creation, execution, failure analysis, and replanning.}

\subsection{Task Creation (Strategist)}

The \SysName workflow commences with a user-defined high-level objective, such as \textit{``obtaining \texttt{user.txt} from \texttt{10.0.2.5}''}. \JY{Upon receiving the objective, the Strategist interprets the goal and constructs an initial task set based on the currently available target information. Subsequent graph expansion driven by execution feedback and newly discovered clues is described in Section \ref{eval:task-replanning}.}

To model relations among tasks, \SysName organizes the overall objective as a directed acyclic graph (DAG), \JY{where nodes represent trackable and schedulable task units, and directed edges encode prerequisite dependencies. Clues are stored separately in the Clue Graph and are associated with their producer and potential consumer tasks through explicit mappings.}\JY{This structure allows the system to preserve multiple potential attack branches and, when one branch becomes invalid, backtrack to a valid upstream node and activate other dependency-satisfied candidates.}
Critically, \SysName maintains a clear distinction between planning granularity and execution-level actions. The Strategist Agent generates semantically complete tasks rather than atomic command-line inputs. Once a task is sufficiently concrete for the Executor Agent to operationalize, planning-level decomposition ceases, and the task is treated as an atomic planning unit. While a single planning unit may translate into multiple consecutive execution steps, the boundary of decomposition is governed by the requirement for high-level reasoning rather than the number of underlying commands. Furthermore, each task is initialized with comprehensive metadata, including task descriptions, prerequisites, relevant clues and expected outputs, to support robust dynamic scheduling and post-mortem analysis.

\subsection{Task and Clue Orchestration}
\begin{wrapfigure}{r}{0.40\columnwidth}
    \centering
    \vspace{-10pt}

    \includegraphics[
        width=\linewidth
    ]{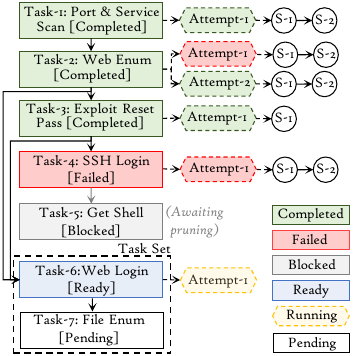}

    \vspace{-0.3\baselineskip}
    \caption{State transition diagram of penetration tasks}
    \label{fig:taskview-1}

    \vspace{-0.4\baselineskip}
\end{wrapfigure}
Upon task creation, \SysName manages the task lifecycle through a dedicated Task \& Clue Cache within the database. This cache serves as the foundational infrastructure for task management and cross-stage reasoning, recording the evolution of task states while accumulating critical clues gathered during execution. For each task, the cache maintains a unique identifier, its topology, dependencies, current state, and associated execution records. As illustrated in Figure~\ref{fig:taskview-1}, task states are strictly categorized into six types: \textit{pending}, \textit{ready}, \textit{running}, \textit{completed}, \textit{failed}, and \textit{blocked}.

Specifically, \textit{pending} denotes a task awaiting its dependencies, whereas \textit{ready} indicates that all execution conditions have been satisfied. \JY{A newly generated task is first registered as \textit{pending}. Deterministic control logic immediately promotes it to \textit{ready} when all prerequisite dependencies have been satisfied.} As shown in Figure~\ref{fig:taskview-1}, ``Task 6: Web Login'' is \textit{ready} because its dependency, ``Task 2: Web Enumeration,'' has been completed. Conversely, ``Task 7: File Enumeration'' remains \textit{pending} until Task 6 is finished.
The \textit{running} state denotes a task currently under execution. To prevent concurrency anomalies caused by simultaneous context writes, \SysName employs a blocking mechanism to ensure that only one task is in the \textit{running} state at any given time. The \textit{completed} and \textit{failed} states are updated based on the outcomes provided by the Executor Agent. To enhance the success rate, the agent may refine and retry failed tasks multiple times. If a task is ultimately deemed a failure, dependent tasks are marked as \textit{blocked}, indicating they have become invalid and have been pruned from the active task graph. For instance, the failure of ``Task 4: SSH Login'' triggers the transition of ``Task 5: Get Shell'' to \textit{blocked}, resulting in its subsequent pruning. While the Reviewer Agent does not manage tasks directly, it provides analytical feedback to the Strategist Agent, which subsequently creates new tasks or updates existing states. Finally, to maintain state consistency and minimize LLM computational overhead, certain transitions, such as the progression from \textit{pending} to \textit{ready}, are handled programmatically rather than via model inference.

Concurrently, the Executor Agent systematically ingests attack-critical telemetry as structured clues. These include open-port fingerprints, DOM structures, configuration files, credential fragments, privilege escalations, error logs, and historical failure data. Rather than naively concatenating this raw data into an LLM prompt, \SysName organizes these fragmented observations into structured evidence that can be explicitly referenced by subsequent tasks.
Through this decoupled design, separating task states from evidentiary clues, \SysName can maintain a persistent shared memory across planning sessions, thereby ensuring the continuity of long-horizon reasoning. Consequently, the Task \& Clue Cache functions as more than a historical record; it actively informs future decision-making, allowing \SysName to transcend the context window constraints inherent in single-turn LLM interactions.

\subsection{Task Execution (Executor)}

Once a task reaches the \textit{ready} state, it could be dispatched to the Executor Agent. The Executor translates the structured task into specific actions by synthesizing its description, prerequisites, and cached clues to dynamically construct a sequence of execution steps. These steps are subsequently mapped to concrete command-line invocations or remote interactions.
The Executor Agent interacts flexibly with the CLI and the underlying operating system, leveraging standard pentesting utilities, such as \texttt{Nmap}, \texttt{Gobuster}, and \texttt{ffuf}, without reliance on predefined Model Context Protocol (MCP) handlers or hardcoded skill sets. This design ensures robust coverage for diverse scenarios while enhancing system flexibility and extensibility. Furthermore, step-level execution traces and session states are recorded as metadata attached to their respective task nodes; \JY{For a single planning-level task, the Executor may perform multiple consecutive command-level attempts, including parameter adjustment, credential substitution, or interaction repair. These attempts remain internal to the task and are aggregated into a single execution trace.} However, these granular logs are not promoted to planning-level nodes to maintain a clean abstraction between low-level execution and high-level strategy.

To address the inherent complexity of \JY{non-linear} pentesting, \SysName supports two distinct execution modes:
1) Non-interactive execution tailored for state-independent operations such as port scanning, directory enumeration, and page retrieval. 2) Interactive execution utilizes a pseudo-terminal (PTY) mechanism to support tasks requiring persistent session states. This includes command chaining following a successful shell exploit, context preservation after remote login, and multi-turn interactions necessitating real-time feedback.
By leveraging the PTY mechanism, the Executor Agent can issue commands and process outputs within a single terminal session while preserving local context --- including the current working directory, user identity, and environment variables. This prevents the fragmentation of context-dependent operations into isolated processes. 
\JY{When the task terminates, the Executor aggregates stdout, stderr, exit codes, network logs, session metadata, and newly observed clues. It then performs a preliminary completion assessment against the task’s expected output. The complete evidence package is written to the cache, while root-cause attribution and cross-task interpretation of unsuccessful or inconclusive outcomes are deferred to the Reviewer.}

\subsection{Result Analysis (Reviewer)}

In automated pentesting, task failure does not necessarily imply that the current attack path is entirely invalid; it may also arise from improper tool usage, incorrect parameter settings, unsatisfied prerequisites, or a failure to correlate critical clues. Therefore, we introduce the Reviewer Agent, an independent component designed to analyze task failures and translate them into structured explanations usable for subsequent decision-making.

\JY{During review, the Reviewer examines the failed task description, its expected output, the complete execution trace, and the contextual clues retrieved from the Task \& Clue Cache. The analysis proceeds in two stages. First, it determines whether the failure is caused by: (1) an incorrect high-level attack path, (2) a localized execution issue involving tool usage, command construction, or interaction flow, or (3) an unsatisfied prerequisite. Second, independently of the failure category, it inspects the execution output for newly exposed evidence, such as usernames, endpoints, file paths, service metadata, or configuration fragments.}

Notably, while the Strategist Agent is designed to encourage broad exploration and divergent heuristic attempts, \JY{the Reviewer Agent helps converge the search space by identifying unviable tasks and recommending branch termination or switching when necessary. If the Reviewer determines that a failure is irrecoverable, the verdict is passed to the Strategist and the programmatic state-control logic. The original failed task and its execution evidence are retained for future reference, while downstream tasks that strictly depend on that node are marked as blocked and removed from the active scheduling set. Completed upstream tasks remain unchanged, and their accumulated clues remain available to other branches.}

\subsection{Task Re-Planning (Strategist)}\label{eval:task-replanning}

\begin{wrapfigure}{r}{0.45\columnwidth}
    \vspace{-\intextsep}
    \centering

    \includegraphics[
        width=\linewidth
    ]{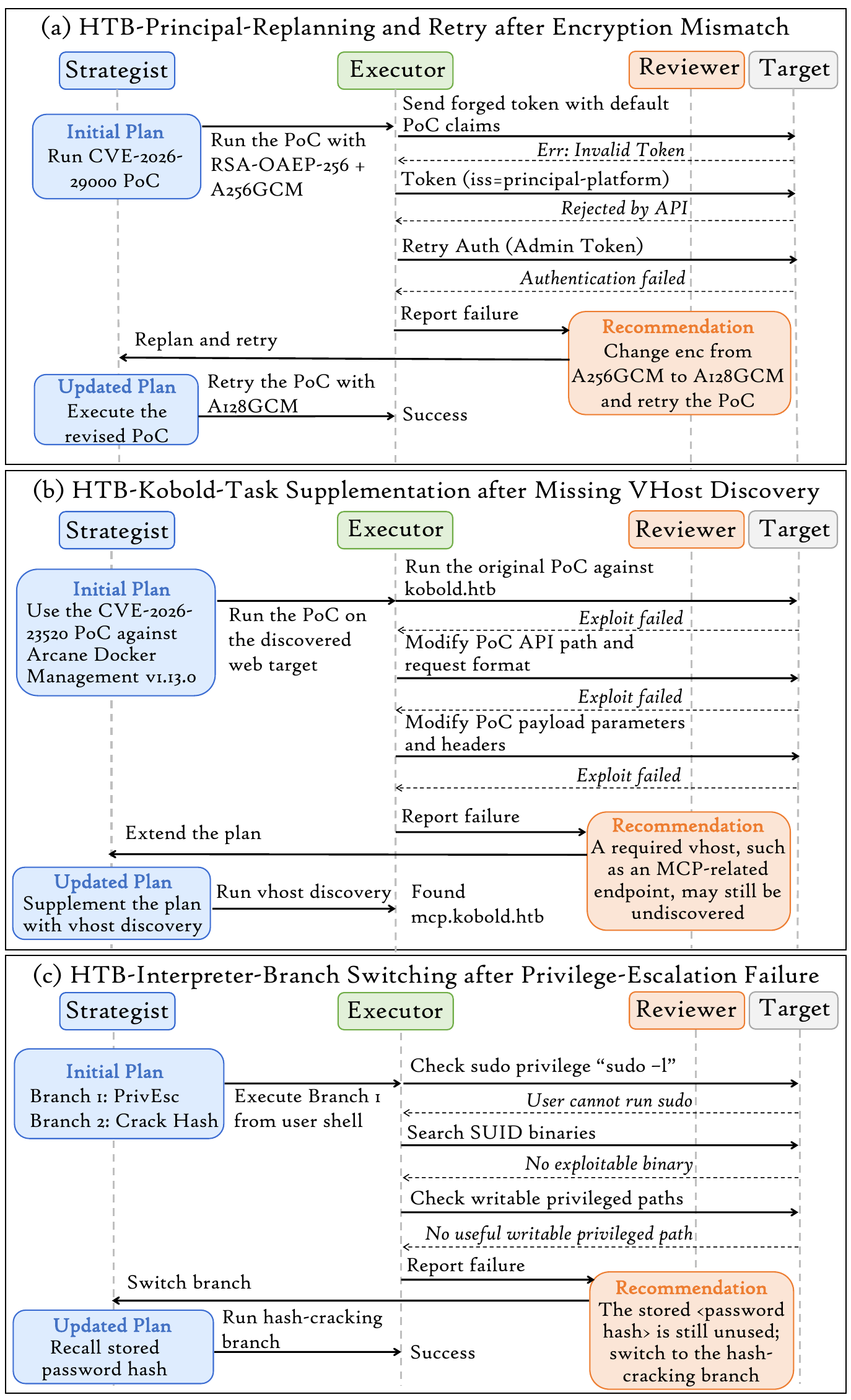}

    \vspace{-0.4\baselineskip}
    \caption{Running examples}
    \label{fig:running-example}
    \vspace{-0.7\baselineskip}
\end{wrapfigure}
Following the review phase, \SysName initiates the replanning stage to generate subsequent tasks. At this point, the system synthesizes the latest task states, dependency structures, failure diagnoses, and newly discovered clues to locally expand and adjust the active task graph, thereby determining the optimal next step. That is, the true autonomy of \SysName lies not merely in automated task execution, but in its capacity to dynamically adapt attack trajectories based on real-time feedback.

Candidate tasks for subsequent execution originate primarily from three sources: (1) local revisions and re-executions of failed tasks with repairable causes; (2) entirely new tasks spawned by newly discovered clues; and (3) alternative branches accessed by backtracking to a valid upstream node after an invalid branch is pruned. Throughout this process, the Strategist Agent dynamically schedules tasks by evaluating dependency satisfaction, the heuristic value of the current path, clue relevance, historical failure context, and the execution state.
Crucially, the recommendations provided by the Reviewer Agent are not strictly negative or punitive; rather, they frequently offer constructive insights that enable the system to perform incremental adjustments around localized failures or newly revealed clues. For instance, following an unsuccessful web exploitation attempt, the system does not prematurely abandon the target. Instead, it generates supplementary tasks --- such as inspecting backup files, testing default credentials, or enumerating API endpoints. Similarly, when the Reviewer Agent extracts novel usernames, file paths, or configuration details from error logs, this new evidence can trigger subsequent authentication attempts, local file reading, or privilege escalation sequences.

\JY{Ultimately, \SysName forms a closed loop in which the Strategist maintains and revises the Task Graph, the Executor interacts with the target and records evidence, the Reviewer diagnoses unsuccessful executions, and the persistent cache preserves tasks and clues across iterations. This loop enables the system to revise local actions, supplement missing prerequisites, and switch branches without discarding valid upstream progress.}

\subsection{Running Examples}
Next, we present three representative running examples, curated from approximately tens of thousands of lines of interaction logs across all target machines, to intuitively illustrate our architectural design choices. These specificscenarios manifest frequently throughout the execution traces.
Because pentesting is fundamentally a black-box process, iterative trial and error is inevitable. For a well-defined task, such as ``\textit{Run the PoC with RSA-OAEP-256 + A256GCM}'', the Executor Agent naturally translates the objective into concrete scripts, systematically iterating over various parameters or credentials. However, LLMs are highly susceptible to a depth-first trap.
Intuitively, as an agent accumulates execution context from repeated failures within a single task, its attention becomes disproportionately anchored to that specific local branch. This hyper-focus inevitably leads to goal drift and the generation of hallucinated, redundant tasks, a critical vulnerability frequently observed in simple ``Planner-Executor'' baselines such as PentestGPT-V2 and Claude Code.

To mitigate this limitation, the Reviewer Agent intervenes at critical junctures to analyze the root causes of failures and furnish actionable recommendations. As illustrated in Figure~\ref{fig:running-example}, these interventions primarily occur in three ways:
1) Local Task Adjustment (Figure~\ref{fig:running-example} (a)): Refining the parameters of the current task, such as correcting the chosen encryption algorithm.
2) Prerequisite Identification (Figure~\ref{fig:running-example} (b)): Diagnosing missing execution conditions and directing the Strategist Agent to formulate the necessary prerequisite tasks.
3) Task Switching (Figure~\ref{fig:running-example} (c)): If local adjustments repeatedly fail, the system leverages the shared {Task \& Clue Cache} to abandon the dead-end branch and proactively pivot to an alternative attack path.
Through this synergistic division of labor, the three agents effectively collaborate to systematically categorize and resolve execution anomalies, thereby accomplishing penetration tasks with significantly enhanced speed and overall efficacy.

\section{Implementation}
\label{sec:implementation}

The implementation and runtime environment of \SysName is architected as a cohesive ecosystem comprising a collaborative multi-agent framework, a specialized execution backend, and a persistent storage layer. At its core, the system utilizes Kimi Code CLI\footnote{Kimi Code, \url{https://www.kimi.com/code/docs/en/}} to orchestrate a Strategist-Executor-Reviewer architecture, which facilitates a natural-language interface analogous to conventional CLI agents for defining attack objectives and environmental constraints. While this architecture supports human-in-the-loop intervention for future research, all evaluations in this study were initiated through a single initial task specification to rigorously assess the system's operational autonomy.

Complementing this reasoning layer, all low-level operations are routed to a Kali Linux-based execution backend to leverage industry-standard utilities such as \texttt{Nmap}, \texttt{Gobuster}, \texttt{ffuf}, \texttt{curl}, \texttt{wget}, \texttt{ssh}, \texttt{ftp}, and \texttt{smbclient}. Within this backend, the Executor Agent employs two distinct shell invocation modes, including a non-interactive mode for rapid scanning and enumeration as well as a PTY-based interactive mode specifically designed for tasks requiring persistent sessions, such as handling authentication prompts or stabilizing reverse shells. To mitigate the blocking behavior common in long-running security tasks, the framework incorporates a dataflow-driven control mechanism that monitors process idle time rather than relying on rigid, fixed timeouts.

Furthermore, to mitigate hallucinations and ensure long-horizon persistence beyond the inherent constraints of the LLM context window, \SysName externalizes the global investigation state into a MySQL-backed persistent storage layer. This storage is partitioned into two specialized, interdependent structures: a Task Graph that maintains the global attack topology, inter-task dependencies, and execution states, and a Clue Graph that serves as a repository for distilled intelligence, such as discovered credentials and vulnerability fragments. Crucially, the Clue Graph maintains explicit mappings between producer tasks, those that generate specific clues, and potential consumer tasks that leverage or derive from those findings, thereby enabling the efficient cross-stage reasoning necessary for complex, non-linear pentesting scenarios.

\section{Evaluation}
\label{sec:evaluation}

In this section, we focus on evaluating the performance of \SysName by answering the following research questions:

\vspace{-\topsep}
\begin{list}{\labelitemi}{\leftmargin=1.5em}
 \setlength{\topmargin}{0pt}
 \setlength{\itemsep}{0em}
 \setlength{\parskip}{0pt}
 \setlength{\parsep}{0pt}

    \item \textbf{RQ1:} How effective is \SysName at autonomously completing end-to-end penetration tasks? (\S\ref{eval:e2e})
    \item \textbf{RQ2:} What are the primary root causes and operational bottlenecks that lead to task failures? (\S\ref{sec:fail_analysis})
    \item \textbf{RQ3:} What is the computational and temporal overhead of \SysName? (\S\ref{eval:overhead})
    \item \textbf{RQ4:} How do the individual architectural components contribute to the overall performance? (\S\ref{eval:orchestration})
\end{list}

\subsection{Evaluation Setup}
\label{subsec:evaluation_methodology}

All experiments were conducted on three dedicated servers running Kali Linux to ensure the seamless invocation of pre-installed pentesting utilities. Regarding LLM access, all models were queried via standard pay-as-you-go (token-based) APIs, with the sole exception of the Claude Code framework powered by Sonnet 4.5, which utilized a flat-rate monthly subscription.

\subsubsection{Baselines, Targets and Foundational Model Selection}
\label{subsubsec:dataset_selection}

The selection of evaluation targets was governed by three primary criteria. First, the dataset must support comprehensive scenario coverage, encompassing diverse operating systems, varying complexity levels, and prevalent Common Weakness Enumerations (CWEs). Second, the targets must necessitate non-linear, multi-step exploitation trajectories to adequately benchmark the systems' advanced reasoning and orchestration capabilities within complex environments. Third, to rigorously maintain evaluation integrity, the publication dates of all selected targets must strictly postdate the training data cutoffs of the evaluated LLMs.

Consequently, we curated a dataset comprising all 10 HTB machines from Season 10 (4 Easy, 4 Medium, and 2 Hard targets) available as of our data collection cutoff in April 2026. These machines were officially released between February 2026 and April 2026. This timeline fully satisfies the knowledge cutoff constraints of all selected LLMs, effectively eliminating the risk of evaluation bias stemming from data contamination, as previously demonstrated in \S\ref{sec:background}.
Regarding baseline systems, we selected PentestGPT-V2, an open-source, specialized framework capable of autonomously executing the end-to-end pentesting lifecycle, as illustrated in Table~\ref{tab:systemComparison}. Furthermore, we incorporated Claude Code, a general-purpose coding agent that has recently demonstrated remarkable efficacy in automated offensive security tasks. Due to compatibility constraints inherent to Claude Code, the aforementioned baseline systems were evaluated using two foundational models: Claude Sonnet 4.5 and DeepSeek-V3.2. In contrast, our proposed \SysName framework underwent extended evaluation utilizing GPT-5.2 and GPT-5.2-Codex, in addition to the models tested with the baselines.



\subsubsection{Experimental Protocol}
\label{subsubsec:experimental_protocol}


pentesting is inherently a complex, open-ended challenge. In real-world engagements, the precise locations of vulnerabilities and exploitable clues, alongside the requisite steps to construct a viable attack path, remain completely unknown prior to execution. While the selected HTB targets possess canonical solution paths, this does not negate the value of exploring alternative, non-standard attack vectors.
On the contrary, if an automated system immediately isolates the correct execution branch and achieves the objective in a single, flawless attempt, it is highly indicative of data contamination (i.e., knowledge leakage from training corpora) rather than genuine reasoning and exploration capabilities. Furthermore, due to the inherent stochasticity of LLM inference, the generated task plans for a given scenario will inevitably vary across different iterations, exhibiting natural fluctuations in both execution sequencing and task granularity.
Consequently, achieving a quantified, strictly objective, and perfectly precise evaluation of pentesting efficacy remains highly challenging, rendering manual review indispensable for data annotation and result aggregation.

Prior works typically evaluate performance based on the binary success rate of obtaining a Flag or Shell~\cite{singer2025incalmo}, an approach that obscures critical nuances. Alternatively, some studies rely on the ``number of completed sub-tasks'' as a metric~\cite{deng2026makes}; however, this overlooks the system's exploration of incorrect yet meaningful attack branches. In this paper, we propose quantifying the absolute number of explored task branches, such as the enumeration of distinct ports or the execution of diverse exploit vectors targeting a single service, to measure the system's capacity for lateral exploration.
Furthermore, existing literature severely lacks in-depth failure analysis. Our empirical results demonstrate that in open-ended environments devoid of knowledge leakage, automated pentesting exhibits substantial stochasticity and a notable probability of failure. The root causes of these failures often stem from specific limitations in LLM capabilities (e.g., information retrieval errors or flawed payload generation) or from highly repetitive tasks (e.g., brute-force password cracking) that are fundamentally ill-suited for LLMs. A rigorous analysis of these failure modes is crucial for understanding current bottlenecks and guiding future research.

Automated penetration testing frameworks may continue generating new tasks even after useful progress has stalled, making natural termination unsuitable for controlled comparison. Wall-clock limits are also unstable because scanning, connection attempts, and network timeouts introduce substantial environmental variance. We therefore use cumulative LLM token consumption as the primary resource constraint and impose a maximum budget of 20 million tokens on each system–target run. Token consumption includes the input and output tokens reported by the corresponding model provider. Because tokenization differs across model families, direct effectiveness comparisons are made primarily between frameworks using the same backbone model.
The 20-million-token threshold provides sufficient budget for sustained exploration while preventing a small number of repetitive runs from dominating the evaluation cost. If a framework terminates naturally before reaching the budget, we retain its last verified penetration state. If a run exceeds the budget, only actions completed within the first 20 million tokens are included in the effectiveness results. Section 5.4 separately analyzes the complete execution traces to characterize the practical resource consumption of sustained execution.

To reduce target-specific solution leakage, all agents are prohibited from accessing machine-specific walkthroughs, write-ups, flag disclosures, or search results that explicitly identify the target’s canonical attack path. Access to general vulnerability databases, vendor advisories, exploit repositories, and documentation is permitted because such resources are available in realistic penetration testing. We manually inspect the browsing and command traces for target-specific leakage. If a framework accesses a prohibited write-up, all actions after that access are discarded, and the last independently verified penetration state before the access is retained for the formal effectiveness analysis. We therefore describe the setting as target-write-up-free rather than zero-knowledge, because the agents retain access to general vulnerability knowledge and public exploit resources.


\definecolor{hiveBlue}{RGB}{237,245,252}
\definecolor{gainGreen}{RGB}{28,130,70}
\definecolor{tieGray}{RGB}{105,105,105}
\definecolor{headerGray}{RGB}{242,242,242}

\newcolumntype{C}{>{\centering\arraybackslash}X}
\newcolumntype{H}{%
  >{\columncolor{hiveBlue}\centering\arraybackslash}X%
}

\newcommand{\gain}[1]{%
  \textcolor{gainGreen}{\textbf{$\uparrow\, +#1$}}%
}

\newcommand{\tie}{%
  \textcolor{tieGray}{$\rightarrow\, 0$}%
}

\newcommand{\ushell}{\textsuperscript{\scriptsize U}}
\newcommand{\rshell}{\textsuperscript{\scriptsize R}}

\begin{table*}[t]
\centering

\caption{
End-to-end progress and shell acquisition under a common maximum budget of 20M LLM tokens per target run
}
\label{tab:all-system-model-results}

\begingroup
\footnotesize
\setlength{\tabcolsep}{2.3pt}

\begin{tabular}{llllllllllllccc}

\toprule


\multirow{2}{*}{Model}
&
\multirow{2}{*}{Framework}
&
\multicolumn{10}{c}{Target-wise Progress $\leq$20M}
&
\multicolumn{3}{c}{Overall Results $\leq$20M}
\\

\cmidrule(lr){3-12}
\cmidrule(l){13-15}


&
&
\makecell{Wing-\\Data}
&
CCTV
&
Kobold
&
\makecell{Silen-\\tium}
&
\makecell{Ptero-\\dactyl}
&
\makecell{Inter-\\preter}
&
\makecell{Princi-\\pal}
&
\makecell{Dev-\\Area}
&
Pirate
&
\makecell{Gar-\\field}
&
\makecell{Completion\\Rate}
&
\makecell{User\\Shell}
&
\makecell{Root\\Shell}
\\

\midrule


\multirow{3}{*}{\makecell[l]{Sonnet 4.5}}

&
\cellcolor{hiveBlue}\SysName
&
\cellcolor{hiveBlue}7/13\ushell
&
\cellcolor{hiveBlue}5/11
&
\cellcolor{hiveBlue}12/12\rshell
&
\cellcolor{hiveBlue}9/14\ushell
&
\cellcolor{hiveBlue}8/14\ushell
&
\cellcolor{hiveBlue}8/14\ushell
&
\cellcolor{hiveBlue}13/13\rshell
&
\cellcolor{hiveBlue}4/15
&
\cellcolor{hiveBlue}2/21
&
\cellcolor{hiveBlue}3/22
&
\cellcolor{hiveBlue}\textbf{47.65\%}
&
\cellcolor{hiveBlue}\textbf{6}
&
\cellcolor{hiveBlue}\textbf{2}
\\

&
Claude Code
&
7/13\ushell
&
4/11
&
5/12
&
7/14
&
5/14
&
8/14\ushell
&
7/13
&
4/15
&
2/21
&
2/22
&
34.23\%
&
2
&
0
\\

&
PentestGPT-V2
&
7/13\ushell
&
4/11
&
5/12
&
7/14
&
5/14
&
5/14
&
12/13\ushell
&
4/15
&
2/21
&
3/22
&
36.24\%
&
2
&
0
\\

\midrule


\multirow{3}{*}{\makecell[l]{DeepSeek-V3.2}}
&
\cellcolor{hiveBlue}\SysName
&
\cellcolor{hiveBlue}7/13\ushell
&
\cellcolor{hiveBlue}3/11
&
\cellcolor{hiveBlue}9/12\ushell
&
\cellcolor{hiveBlue}8/14
&
\cellcolor{hiveBlue}9/14\ushell
&
\cellcolor{hiveBlue}3/14
&
\cellcolor{hiveBlue}13/13\rshell
&
\cellcolor{hiveBlue}4/15
&
\cellcolor{hiveBlue}2/21
&
\cellcolor{hiveBlue}1/22
&
\cellcolor{hiveBlue}\textbf{39.60\%}
&
\cellcolor{hiveBlue}\textbf{4}
&
\cellcolor{hiveBlue}\textbf{1}
\\

&
Claude Code
&
7/13\ushell
&
4/11
&
7/12\ushell
&
8/14
&
5/14
&
4/14
&
7/13
&
4/15
&
2/21
&
1/22
&
32.89\%
&
2
&
0
\\

&
PentestGPT-V2
&
7/13\ushell
&
3/11
&
5/12
&
8/14
&
9/14\ushell
&
3/14
&
13/13\rshell
&
4/15
&
2/21
&
1/22
&
36.91\%
&
3
&
1
\\

\midrule


GPT-5.2
&
\cellcolor{hiveBlue}\SysName
&
\cellcolor{hiveBlue}7/13\ushell
&
\cellcolor{hiveBlue}5/11
&
\cellcolor{hiveBlue}7/12\ushell
&
\cellcolor{hiveBlue}10/14\ushell
&
\cellcolor{hiveBlue}5/14
&
\cellcolor{hiveBlue}8/14\ushell
&
\cellcolor{hiveBlue}13/13\rshell
&
\cellcolor{hiveBlue}9/15
&
\cellcolor{hiveBlue}4/21
&
\cellcolor{hiveBlue}4/22
&
\cellcolor{hiveBlue}48.32\%
&
\cellcolor{hiveBlue}5
&
\cellcolor{hiveBlue}1
\\

\midrule


\makecell[l]{GPT-5.2-Codex}
&
\cellcolor{hiveBlue}\SysName
&
\cellcolor{hiveBlue}7/13\ushell
&
\cellcolor{hiveBlue}5/11
&
\cellcolor{hiveBlue}7/12\ushell
&
\cellcolor{hiveBlue}9/14\ushell
&
\cellcolor{hiveBlue}5/14
&
\cellcolor{hiveBlue}8/14\ushell
&
\cellcolor{hiveBlue}13/13\rshell
&
\cellcolor{hiveBlue}8/15
&
\cellcolor{hiveBlue}4/21
&
\cellcolor{hiveBlue}4/22
&
\cellcolor{hiveBlue}46.98\%
&
\cellcolor{hiveBlue}5
&
\cellcolor{hiveBlue}1
\\

\bottomrule

\end{tabular}

\endgroup

\end{table*}

\subsection{End-to-End Effectiveness}\label{eval:e2e}

First, we evaluate the end-to-end task completion capabilities of various system-LLM combinations (combination for short) based on the volume of resolved sub-tasks and the successful acquisition of user and root shells. As summarized in Table~\ref{tab:all-system-model-results}, \JYY{we report the completed and total number of sub-tasks for each target, with ground-truth task counts manually established by synthesizing target specifications with the task-partitioning logic of various automated frameworks (refer to Appendix Table~\ref{tab:gt-inter} for a representative example). Superscripts U and R denote the acquisition of user and root shells, respectively.}
The results demonstrate that \SysName achieves higher capped completion rates with both shared backbone models. Specifically, when utilizing the Claude Sonnet 4.5 model, \SysName successfully secures \JYY{two} additional root shells and \JYY{four} additional user shells \JYY{compared with both Claude Code and PentestGPT-V2.} We observe comparable performance advantages when evaluating the systems with the DeepSeek-V3.2 model. It is critical to note that Claude Code and its derivative, PentestGPT-V2, employ a dynamic sub-agent mechanism. These sub-agents occasionally bypassed our experimental constraints regarding the use of online write-ups; we identified clear evidence of knowledge leakage in at least two instances. In contrast, \SysName maintains strict adherence to zero-leakage constraints. When these contaminated cases are excluded, the performance advantage of our proposed framework becomes even more pronounced.

To evaluate lateral exploration, we define ``Exploration Breadth'' as $\sum_{v\in V}\max(\operatorname{outdeg}(v)-1,0)$ across the pentesting decision graph. This metric explicitly quantifies a system's capacity for multi-branch decision-making: rigid, linear execution (e.g., A $\rightarrow$ B $\rightarrow$ C) scores 0, whereas spawning parallel attack vectors from a single pivot point (e.g., simultaneous directory fuzzing and SSH brute-forcing) positively increases the score. As shown in Figure~\ref{fig:exploration_breadth}, \SysName outperforms both baselines, achieving the highest median breadth and demonstrating highly adaptive scaling on complex targets. In contrast, Claude Code's consistently low scores quantitatively reflect its tendency for premature abandonment, while PentestGPT-V2's narrow interquartile range indicates a reliance on redundant, single-threaded execution loops rather than strategic lateral expansion.
Crucially, this robust multi-branch exploration directly facilitates greater penetration depth. The aforementioned results were observed within the relatively constrained bounds of experimental target machines; this capacity for extensive lateral exploration becomes indispensable when navigating the complexities of real-world enterprise environments.

\subsection{Failure Reason Analysis}
\label{sec:fail_analysis}

\begin{figure}[!t]
\centering
\vspace{-0.05in}

\begin{minipage}[t]{0.34\linewidth}
    \vspace{0pt}
    \centering

    \includegraphics[
        width=0.99\linewidth
    ]{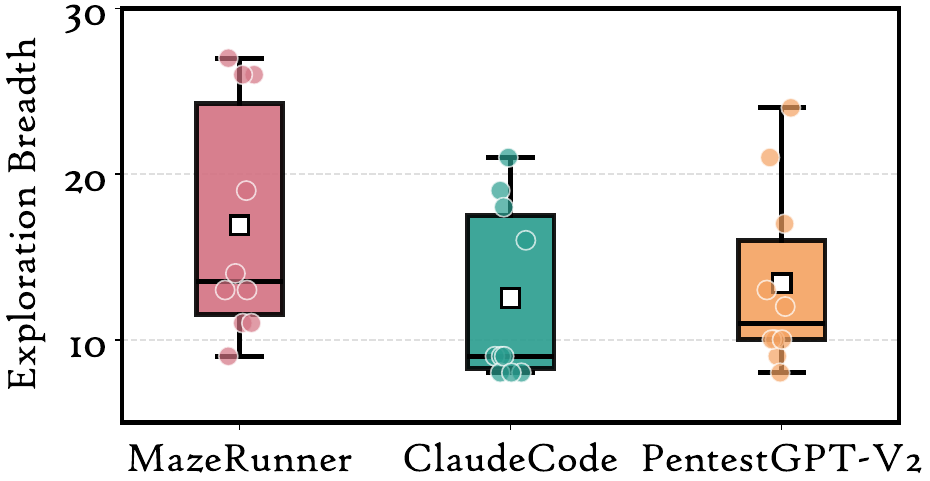}

    \vspace{-0.03in}
    \captionof{figure}{
        Exploration breadth by system
    }
    \label{fig:exploration_breadth}
\end{minipage}%
\hfill
\begin{minipage}[t]{0.64\linewidth}
    \vspace{0pt}
    \centering

    \captionof{table}{
        Distribution of failure causes across frameworks and backbone LLMs
    }
    \label{tab:error_analysis}

    \vspace{-0.01in}

    \begingroup
    \scriptsize
    \setlength{\tabcolsep}{1.8pt}
    \renewcommand{\arraystretch}{1.08}

    \begin{tabular}{lcccccc}
        \toprule
        &
        \makecell[c]{Unsuccessful\\Traces}
        &
        \makecell[c]{Planning\\Error}
        &
        \makecell[c]{Clue\\Forgetting}
        &
        Abandonment
        &
        \makecell[c]{Execution\\Failure}
        &
        \makecell[c]{Retrieval\\Failure}
        \\
        \midrule

        \multicolumn{7}{l}{
            \textit{Panel A: By Framework}
        }
        \\[-1pt]

        \SysName
        & 17
        & 29.4\%
        & 5.9\%
        & 23.5\%
        & 52.9\%
        & 5.9\%
        \\

        Claude Code
        & 20
        & 55.0\%
        & 25.0\%
        & 30.0\%
        & 25.0\%
        & 5.0\%
        \\

        PentestGPT-V2
        & 19
        & 68.4\%
        & 21.1\%
        & 10.5\%
        & 36.8\%
        & 0.0\%
        \\

        \midrule

        \multicolumn{7}{l}{
            \textit{Panel B: By Backbone LLM}
        }
        \\[-1pt]

        Sonnet 4.5
        & 28
        & 50.0\%
        & 21.4\%
        & 32.1\%
        & 25.0\%
        & 0.0\%
        \\

        DeepSeek-V3.2
        & 28
        & 53.6\%
        & 14.3\%
        & 10.7\%
        & 50.0\%
        & 7.1\%
        \\

        \bottomrule
    \end{tabular}

    \endgroup
\end{minipage}

\vspace{-0.08in}
\end{figure}

We manually analyze the 56 target-level runs that do not obtain root access across the three frameworks, two shared backbone models, and 10 targets. Each run is assigned one or more failure categories: Planning Error, Clue Forgetting, Premature Abandonment, Code/Payload Failure, and Information Retrieval Failure. Because failure causes may compound within a run, the categories are treated as non-exclusive. Definitions and representative cases are provided in~\ref{app:fail}. 
As shown in Table~\ref{tab:error_analysis}, MazeRunner exhibits lower observed rates of Planning Error and Clue Forgetting than both baseline frameworks. At the same time, Code/Payload Failures account for a larger proportion of its unsuccessful runs. One possible explanation is that improved task orchestration allows more MazeRunner runs to progress beyond reconnaissance and planning into execution-intensive stages, where exploit compatibility and payload correctness become the dominant bottlenecks. This interpretation is based on the observed traces and should not be treated as a controlled causal result.

Comparing the foundational models reveals that Claude Sonnet 4.5 \JYY{exhibits lower observed rates of code/payload execution and information retrieval failures than DeepSeek-V3.2. This complements MazeRunner's orchestration capability: while the framework mitigates early-stage planning and state-management bottlenecks, Sonnet provides stronger support for the execution-intensive stages that follow.} Conversely, baseline systems frequently stall in the early exploration phases, rendering Sonnet's coding prowess underutilized. Finally, we highlight a notable anomaly regarding GPT-5.2-Codex, a model explicitly fine-tuned for coding tasks. Surprisingly, it underperformed compared to the standard GPT-5.2 model. \JYY{We hypothesize that its built-in task-orchestration behavior interferes with MazeRunner's external orchestration, introducing additional token overhead without commensurate performance gains.}

Furthermore, among the remaining instances of planning failure within our system, the inherent stochasticity of LLM orchestration plays a substantial role. Because LLMs are fundamentally probabilistic, this randomness is unavoidable. In our specific pentesting context, it primarily manifests as a statistical bias during task prioritization: when multiple viable attack vectors exist, the model gravitates toward the most common solutions. For simple tasks or scenarios where the correct exploit aligns with historically frequent vulnerabilities, this bias has minimal impact, as the system can reliably iterate through feasible options utilizing the cached task list.
However, in complex scenarios, particularly when the correct attack vector is obscure or statistically rare, this prioritization bias severely degrades problem-solving efficiency. For instance, during the evaluation of the ``HTB-Garfield'' target, \SysName failed to access the critical ``ACL/scriptPath'' chain. Instead, it disproportionately expended resources executing prevalent, zero-prerequisite attacks, such as Zerologon, administrator credential stuffing, PetitPotam, Kerberoasting, and RDP brute-forcing. Consequently, it entirely overlooked the correct, yet unconventional, strategy of exploiting a Kerberos initial account backdoor. Fundamentally, this dilemma represents a classic trade-off between penetration capabilities and computational overhead. While exhaustively enumerating all obscure attack paths might eventually yield success, it requires a prohibitive expenditure of resources. Moreover, such exhaustive probing would inevitably trigger context window saturation, transforming the LLM's finite memory into a new operational bottleneck.

Finally, other observed execution failures stem from the system attempting highly repetitive, token-intensive tasks, such as autonomously generating scripts for password brute-forcing. To optimize both efficiency and token economy, future iterations of \SysName should explicitly delegate these deterministic, brute-force operations to specialized external tools rather than relying on LLM-driven code generation.

\subsection{Overhead}
\label{eval:overhead}
\begin{figure}[tbp]
  \centering
  \includegraphics[width=0.80\linewidth]{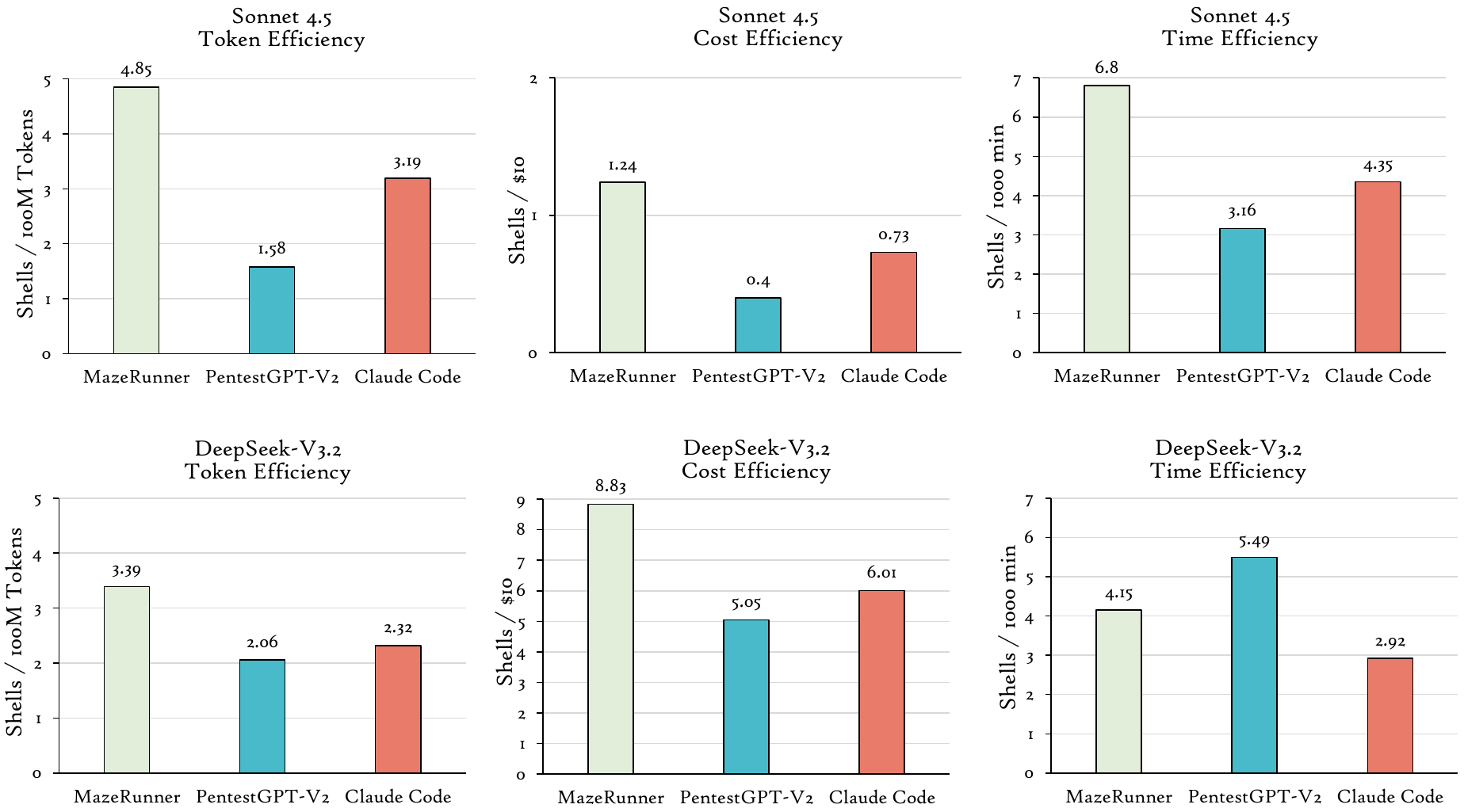}
  \caption{Token, cost, and time efficiency over complete execution traces}
  \label{fig:token-cost}
\end{figure}

\JYY{While the effectiveness analysis in Section~\ref{eval:e2e} is restricted to a common 20M-token budget for fair comparison, the overhead analysis uses the complete execution traces. These two settings provide complementary perspectives: the former compares framework effectiveness under the same resource budget, while the latter characterizes practical resource consumption and efficiency during sustained pentesting. As illustrated in Figure~\ref{fig:token-cost}, we measure shell-acquisition efficiency as the number of successful shell acquisitions per 100M tokens, per \$10, and per 1000 minutes of execution time. \SysName achieves the highest token and cost efficiency under both Claude Sonnet 4.5 and DeepSeek-V3.2, while its time efficiency varies across models. The differences in financial overhead are also affected by substantial disparities in model pricing; the unit cost of Claude Sonnet 4.5 is over tenfold that of DeepSeek-V3.2. Consequently, this leads to a substantial gap in final financial overhead. Overall, DeepSeek-V3.2 achieves competitive penetration performance while maintaining high cost-efficiency.}

Comparing \SysName against the baseline frameworks further demonstrates its operational efficiency. Claude Code exhibits a tendency for premature task termination; while this results in lower overall overhead, it severely degrades its sub-task completion rate. Conversely, PentestGPT-V2, much like our system, engages in continuous exploration. However, owing to its propensity for generating redundant operations and hallucinatory task loops, PentestGPT-V2 consumes approximately twice the tokens and financial budget before its progress stagnates, yet ultimately achieves less penetration depth than \SysName (detailed command-level execution statistics are provided in ~\ref{app:commands}). Finally, temporal expenditure across all combinations exhibits high variance and stochasticity. This randomness is primarily because many pentesting procedures (e.g., port scanning and connection attempts) involve inherent network timeouts and wait periods, which account for a substantial portion of the total execution time. Given the varying depths and breadths of exploration across different tasks, pure execution time provides limited comparative benchmarking value in this context and should therefore be interpreted cautiously.

Furthermore, we analyzed the token consumption distribution across the system's three primary agents. As depicted in Figure~\ref{fig:average-cost}, the Executor Agent accounts for the overwhelming majority, ranging from 75\% to 95\%, of the total token expenditure. The bulk of computational resources is dedicated to low-level payload and script generation, alongside the subsequent parsing of environmental feedback. Executing irrelevant or hallucinatory tasks wastes substantial token budgets without yielding any strategic benefit. Consequently, investing a relatively minor fraction of the token budget in the Strategist and Reviewer agents for high-level planning and proactive error correction helps prevent costly execution loops and reduce unnecessary computational expenditure, ultimately driving down the overall financial and computational overhead of the pentesting lifecycle.

\subsection{Effectiveness of Individual Components}\label{eval:orchestration}

\begin{wrapfigure}{r}{0.42\textwidth}
    \centering
    \vspace{-24pt}

    \includegraphics[
        width=0.98\linewidth
    ]{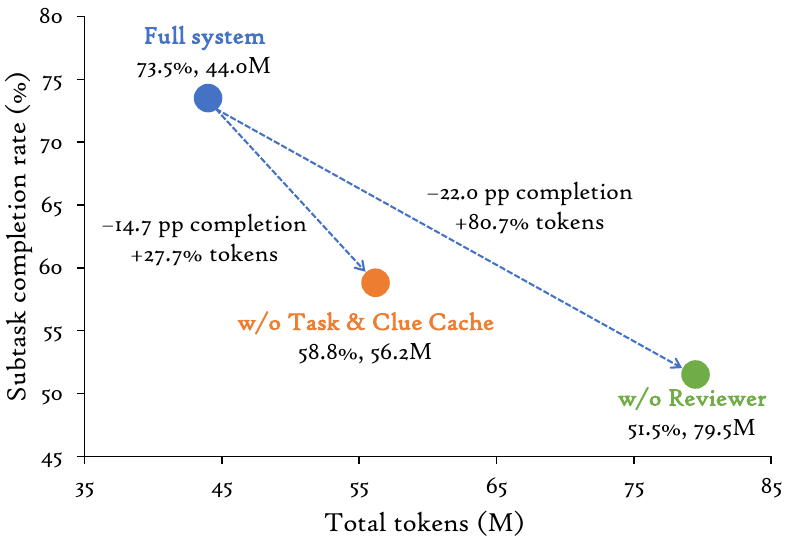}

    \vspace{-0.5\baselineskip}
    \caption{Ablation study results}
    \label{fig:xiaorong}
\end{wrapfigure}

We conducted an ablation study on a subset of 5 targets, where the framework successfully secured at least a user shell utilizing GPT-5.2, to empirically evaluate the individual contributions of \SysName's core modules. Our proposed framework relies on three primary agents, the Strategist, Executor, and Reviewer, augmented by an independent Task \& Clue Cache. Because the Strategist and Executor are indispensable for executing the fundamental pentesting loop, our ablation experiments specifically isolate and remove the Reviewer Agent and the independent cache to quantify their respective impacts.

Removing the independent cache drops the completion rate (defined as the ratio of successfully resolved sub-tasks to total sub-tasks across all 5 target machines) to 58.8\% and inflates overhead by roughly 27\%. This illustrates``Contextual Amnesia'': without persistent memory, agents overwrite critical historical intelligence as their context windows fill, resulting in the redundant re-exploration of previously scanned environments.
Ablating the Reviewer causes the most severe performance collapse. The completion rate plunges from 73.5\% to 51.5\%, while token consumption surges by over 80\% (to 79.5M). This directly corroborates the ``depth-first trap'' phenomenon: without proactive error correction, the system cannot recover from failures, forcing the Executor to blindly waste resources on dead-end paths and hallucinated tasks.
All in all, this study confirms that while the Strategist and Executor drive the attack's momentum, the Reviewer and Cache are indispensable for maintaining trajectory, recovering from errors, and ensuring cost-efficiency.

\section{Related Work}

Planning is a fundamental prerequisite for automated pentesting, requiring agents to continuously evaluate attack vectors and recalibrate paths under severe uncertainty~\cite{xi2025rise}. Existing approaches to attack orchestration~\cite{skandylas2025automated} broadly bifurcate into two primary paradigms: traditional deterministic planning and LLM-driven planning.
The deterministic paradigm encompasses classical planning~\cite{de2024chainreactor}, attack graphs~\cite{obes2013attack}, POMDPs~\cite{schwartz2020pomdp+}, and reinforcement learning~\cite{chen2023gail}. While these methods excel in logical consistency by explicitly modeling states and transitions, their reliance on rigidly structured representations and idealized environments renders them inherently brittle against real-world, black-box targets.

Conversely, LLM-driven planning offers a highly adaptable alternative. Instead of requiring pre-computed states, LLMs dynamically decompose high-level objectives into executable sub-tasks via natural language reasoning and intermediate feedback. Recent scholarship has extensively explored varied reasoning structures, including one-shot~\cite{zhou2022least}, step-wise~\cite{wei2022chain}, hierarchical~\cite{lin2023swiftsage}, tree-structured~\cite{yao2023tree,deng2024pentestgpt}, and reflective planning~\cite{madaan2023self}, providing a compelling foundation for autonomous security decision-making.

Despite these advancements, current LLM planners exhibit critical deficiencies in live pentesting. Unlike generalized domains, real-world targets feature profound black-box uncertainty, deceptive ``rabbit holes,'' and fragmented cross-stage clues. Consequently, effective planning necessitates robust long-term state persistence, the capacity to synthesize chronologically distant evidence, and the agility to backtrack from errors. While both paradigms offer foundational insights, they remain fundamentally inadequate for addressing the core challenge resolved in this work: resilient, long-horizon attack orchestration and error-aware backtracking within highly uncertain environments.

\section{Discussion \& Conclusion}

\subsection{Limitations \& Future Work} 

Despite the advanced orchestration and memory capabilities introduced by \SysName, there are certain pentesting scenarios that cannot achieve full automation. First, a major challenge arises from limitations in multimodal information processing; specifically, the current system is unequipped to handle tasks requiring Graphical User Interface (GUI) interactions. \JYY{Future work could integrate multimodal perception and GUI interaction capabilities to support browser- and desktop-based security operations.} Second, defining strict security boundaries remains a practical hurdle. To avoid disrupting normal business operations during live testing, potentially destructive or intrusive actions still necessitate human expert review. Third, because this research focuses primarily on task orchestration rather than vulnerability discovery, the system struggles to progress in the absence of known vulnerability disclosures or readily available exploit code. 


Furthermore, because LLMs are inherently probabilistic models, they exhibit a statistical bias toward common attack methodologies when presented with multiple potential exploitation paths. Consequently, in scenarios that necessitate rare or unconventional attack vectors, the system may expend significant iterations before attempting the correct method, or worse, prematurely abandon the active clue to pursue alternative branches. Fundamentally, this issue reduces to the classic search-space dilemma of balancing penetration depth against exploration breadth, a challenge that lacks a definitive algorithmic resolution and frequently confounds even expert human analysts. Future work should improve failure diagnosis, recovery, and long-horizon context management. Although the Reviewer corrects many execution failures, ambiguous tool outputs, network timeouts, and repeated exploit failures may still trigger incorrect task transitions. More structured error representations, adaptive retry policies, and rollback mechanisms could improve robustness. Historical trajectory retrieval and experience-based adaptation may further help the system prioritize less common but viable attack paths. In addition, context compression, clue deduplication, and relevance-aware memory retrieval could reduce prompt noise and preserve salient evidence. Visualizing the evolving task graph and clue dependencies may also improve interpretability and facilitate human supervision.

Recent agentic model Kimi K3 with Kimi Code reveals an important shift in penetration success rate by identifing exploitable behaviors or attack paths that are not included in canonical solution paths. These observations suggest that sufficiently capable models with long-horizon reasoning, code generation, and tool-use abilities can sometimes compensate for the absence of elaborate external task planning by continuously interpreting environmental feedback and revising their actions. Nevertheless, such capability does not eliminate the need for explicit orchestration. Its success often depends on large and highly variable token and command budgets, while unconstrained exploration may reduce cost predictability, execution reproducibility, state auditability, and operational safety. MazeRunner is therefore complementary to frontier agents rather than a replacement for them. Its Task \& Clue Cache and independent review mechanism provide a model-agnostic control layer for persistent state management, failure attribution, branch switching, and resource governance. Kimi K3 and Kimi Code can, in turn, serve as a stronger reasoning and execution backbone within MazeRunner, combining their ability to discover non-predefined attack paths with explicit state consistency, controllable execution, and traceable decision making.


\subsection{Conclusion} 

This paper addresses a central challenge of black-box automated
pentesting: the attack graph must be discovered online, multiple
plausible branches compete for a limited execution budget, failures
require attribution rather than blind retry, and decisive clues must
persist across long action horizons. \SysName meets these requirements
through an incremental Strategist-Executor-Reviewer loop and a
persistent Task \& Clue Cache, enabling dynamic graph expansion,
error-aware correction, branch backtracking, and cross-stage evidence
reuse. Across 10 recently released HTB targets, \SysName achieves
higher shell acquisition rates, over 26\% greater aggregate exploration
breadth than the baselines, and substantially lower computational
overhead. These results establish \SysName as a robust and cost-effective
foundation for automated security auditing in open-ended environments.

\Acknowledgements{
This work was supported by the National Natural Science Foundation of China (Grant No. 62402419), the “Pioneer and Leading Goose” R\&D Program of Zhejiang (Grant No. 2025C02263), the Natural Science Foundation of Ningbo (Grant No. 2025J027), the Ningbo Yongjiang Talent Programme, the CCF-Tencent Rhino-Bird Open Research Fund, and the National Key Research \& Development Project of China (Grant No. 2023YFB3106800).
}

\bibliographystyle{scis}
\bibliography{refs}

\clearpage

\appendix
\clearpage

\section{More Evaluation Setup Information and Results}

\subsection{Targets and Ground Truth}
\label{app:htb_description}

\begin{table*}[!b]
    \centering
    \caption{Overview of the selected evaluation dataset and ground-truth}
    \footnotesize
    \label{tab:Evaluation_Dataset}
    
    \resizebox{\linewidth}{!}{%
    \begin{tabular}{c|c|c|c|c|p{4.2cm}|c|p{4.9cm}}\toprule
        {ID} & {Box Name} & {OS} & {Difficulty} & \begin{tabular}[c]{@{}c@{}}{Released}\\ {Date}\end{tabular} & {Problem Domain} & {\makecell{\# of\\subtasks}} & {Brief Description} \\\midrule
        
        1 & WingData & Linux & Easy & 2026/2/14 & Web, Wing FTP Server RCE, Credential Recovery, Tarfile PrivEsc & 13  & Wing FTP (CVE-2025-47812) foothold, credential recovery, and privesc via vulnerable Python tar script. \\
        
        2 & CCTV & Linux & Easy & 2026/3/7 & Web, IoT/CCTV, ZoneMinder SQLi, motionEye PrivEsc & 11  & ZoneMinder SQLi (CVE-2024-51482) for SSH access; privesc via root-running motionEye interface. \\
        
        3 & Kobold & Linux & Easy & 2026/3/21 & AI Tooling, MCP Inspector RCE, Docker Escape & 12 & MCP Inspector unauthenticated RCE foothold; Docker group abuse for container escape. \\
        
        4 & Silentium & Linux & Easy & 2026/4/11 & AI Workflow Platform, Flowise ATO/RCE, Credential Leakage, Gogs RCE & 14  & Flowise ATO/RCE and Docker credential leak, leading to root via local Gogs RCE. \\
        
        5 & Pterodactyl & Linux & Medium & 2026/2/7 & Web, Pterodactyl Panel, Path Traversal/LFI, PEAR-assisted RCE & 14  & Pterodactyl Panel LFI (CVE-2025-49132) chained with PHP PEAR for RCE. \\
        
        6 & Interpreter & Linux & Medium & 2026/2/21 & Web, Mirth Connect RCE, SSTI/Python eval, Internal Service Abuse & 14  & Mirth Connect (CVE-2023-43208) foothold; root RCE via internal XML service and Python SSTI. \\
        
        7 & Principal & Linux & Medium & 2026/3/12 & Cryptographic Trust, JWT/JWE Auth Bypass, SSH CA Misconfig & 13  & JWT auth bypass (CVE-2026-29000) for admin tokens; SSH CA misconfig for root certificate forgery. \\
        
        8 & DevArea & Linux & Medium & 2026/3/28 & DevOps, Apache CXF LFI, Hoverfly API RCE, Sudo/Binary PrivEsc & 15  & Apache CXF LFI leading to Hoverfly API RCE; root via writable binary and sudo abuse. \\
        
        9 & Pirate & Windows & Hard & 2026/3/28 & Active Directory, gMSA Abuse, Kerberoasting, NTLM Relay & 21  & AD chain: BloodHound, gMSA abuse, NTLM relay, and SPN hijacking to compromise DC. \\
        
        10 & Garfield & Windows & Hard & 2026/4/4 & Active Directory, ACL Abuse, RODC, RBCD, Ticket Forgery & 22  & AD escalation via ACL abuse, lateral movement, RBCD against RODC, and Golden Ticket attacks. \\\bottomrule
    \end{tabular}
    }
\end{table*}

Table~\ref{tab:Evaluation_Dataset} details the 10 HTB targets utilized for evaluation in this paper. To mitigate the risk of data contamination (i.e., knowledge leakage from training corpora), we exclusively selected recently released machines. Furthermore, Table~\ref{tab:gt-inter} delineates the sub-tasks and corresponding environmental clues along the primary attack chain, demonstrating that task progression is clue-driven.

\subsection{Cost of \SysName's Agents}
Figure~\ref{fig:average-cost} illustrates the computational cost distribution among \SysName's three core agents across various foundational LLMs.

\subsection{Command Execution Efficiency and Validity}
\label{app:commands}

To rigorously investigate the operational behaviors of the evaluated systems, we first systematically define the taxonomy of an ``invalid command.'' Specifically, a command is classified as invalid if it exhibits any of the following characteristics: 1) {redundant repetition} (e.g., cyclical scanning or homogeneous parameter brute-forcing); 2) {logical contradiction} (e.g., persistently invoking \texttt{sudo} after confirmed privilege denial); 3) {persistent exploit failure} (e.g., continuing to mutate and execute payloads after three consecutive failures); 4) {contextual amnesia} (e.g., re-initiating reconnaissance on previously discovered assets); 5) {intuition deficits} (e.g., attempting blind file writes without verifying permissions, or scanning out-of-scope subnets); 6) {explicit tool execution errors}; 7) {severe trajectory deviation} from the ground-truth attack chain; or 8) {execution hallucinations} (e.g., proceeding under the false assumption of task completion without affirmative environmental feedback).

\begin{figure}[!t]
\centering
\vspace{-0.05in}

\begin{minipage}[t]{0.58\linewidth}
    \vspace{0pt}
    \centering

    \captionof{table}{Statistics of invalid/total commands}
    \label{tab:command_count}

    \vspace{-0.05in}

    \fontsize{6.5pt}{7.5pt}\selectfont
    \setlength{\tabcolsep}{2.2pt}
    \renewcommand{\arraystretch}{1.22}
    
    \resizebox{0.96\linewidth}{!}{%
    \begin{tabular}{l|cc|cc|cc}
        \toprule

        \multirow{2}{*}{ID}
        &
        \multicolumn{2}{c|}{\SysName}
        &
        \multicolumn{2}{c|}{Claude Code}
        &
        \multicolumn{2}{c}{PentestGPT-V2}
        \\

        \cmidrule(lr){2-3}
        \cmidrule(lr){4-5}
        \cmidrule(lr){6-7}

        &
        DeepSeek
        &
        Sonnet
        &
        DeepSeek
        &
        Sonnet
        &
        DeepSeek
        &
        Sonnet
        \\

        \midrule

        WingData
        &
        \cellcolor{orange!25}86/165
        &
        \cellcolor{orange!25}43/155
        &
        \cellcolor{orange!25}64/210
        &
        \cellcolor{orange!25}69/132
        &
        \cellcolor{orange!25}90/194
        &
        \cellcolor{orange!25}48/148
        \\

        CCTV
        &
        29/132
        &
        48/123
        &
        32/119
        &
        42/147
        &
        87/218
        &
        67/211
        \\

        Kobold
        &
        \cellcolor{orange!25}23/133
        &
        \cellcolor{red!25}13/125
        &
        \cellcolor{orange!25}71/222
        &
        55/287
        &
        73/231
        &
        209/578
        \\

        Silentium
        &
        39/205
        &
        \cellcolor{orange!25}28/163
        &
        18/123
        &
        36/125
        &
        19/150
        &
        86/386
        \\

        Pterodactyl
        &
        \cellcolor{orange!25}48/237
        &
        \cellcolor{orange!25}45/132
        &
        35/130
        &
        26/143
        &
        \cellcolor{orange!25}159/486
        &
        121/473
        \\

        Interpreter
        &
        27/95
        &
        \cellcolor{orange!25}42/175
        &
        22/126
        &
        \cellcolor{orange!25}25/135
        &
        86/308
        &
        73/242
        \\

        Principal
        &
        \cellcolor{red!25}80/351
        &
        \cellcolor{red!25}7/79
        &
        22/82
        &
        19/108
        &
        \cellcolor{red!25}28/86
        &
        \cellcolor{orange!25}49/303
        \\

        DevArea
        &
        12/128
        &
        28/210
        &
        49/210
        &
        35/177
        &
        71/333
        &
        58/345
        \\

        Pirate
        &
        62/178
        &
        214/459
        &
        174/452
        &
        119/238
        &
        348/526
        &
        281/390
        \\

        Garfield
        &
        60/100
        &
        98/188
        &
        125/216
        &
        136/244
        &
        129/205
        &
        94/167
        \\

        \bottomrule
    \end{tabular}%
    }

\end{minipage}
\hfill
\begin{minipage}[t]{0.40\linewidth}
    \vspace{0pt}
    \centering

    \includegraphics[
        width=\linewidth,
        height=4.8cm,
        keepaspectratio,
        trim=2mm 4mm 2mm 3mm,
        clip
    ]{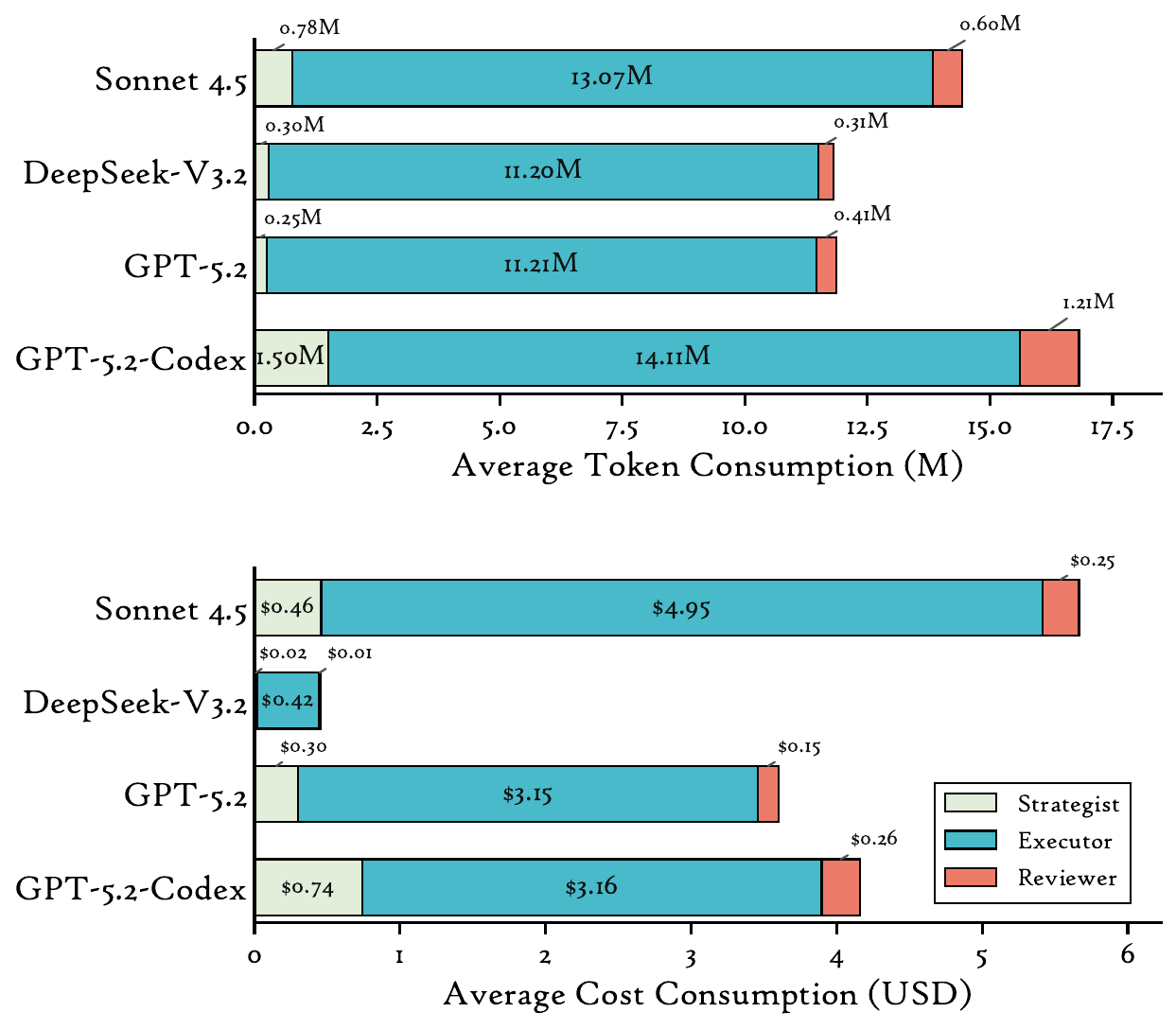}

    \vspace{-0.08in}

    \captionof{figure}{Average cost of \SysName's agents}
    \label{fig:average-cost}

\end{minipage}

\vspace{-0.08in}
\end{figure}

Based on this taxonomy, Table~\ref{tab:command_count} details the ratio of invalid commands to total executed operations across the 10 target machines (where orange and red highlights denote the successful acquisition of User and Root shells, respectively). This fine-grained, command-level data empirically corroborates our prior observations regarding systemic execution bottlenecks. Notably, PentestGPT-V2 exhibits severe execution redundancy. Across complex targets such as Kobold and Pirate, it generates an overwhelming volume of commands (frequently exceeding 400 operations) coupled with a disproportionately high invalid command ratio (e.g., 281/390, or 72\% invalid, on Pirate using Sonnet). This empirically confirms its susceptibility to the ``depth-first trap,'' wherein the agent blindly iterates over hallucinated or flawed payloads without strategic course correction. Conversely, while Claude Code generally issues far fewer total commands, it also secures fewer shells, quantitatively substantiating its tendency toward premature abandonment when confronting execution roadblocks.

\begin{table*}[t]
\centering
\caption{Sub-tasks and clues along the primary penetration path of  ``HTB-Interpreter''}
\label{tab:gt-inter}
\footnotesize
\begin{tabular}{c|l|c|l}\toprule
\# & Task & \makecell{Driven\\By} & Clues \\\midrule
1 & Port Scanning \& Service Identification & $\leftarrow$ & IP Address \\
2 & Web App Fingerprinting \& Version Discovery & $\leftarrow$ & Port 80/443 \\
3 & Vulnerability Intelligence \& Attack Surface Confirmation & $\leftarrow$ & Mirth Connect 4.4.0 \\
4 & Exploit Known Vulnerability for RCE & $\leftarrow$ & CVE-2023-43208 /api/users Deserialization Chain \\
5 & {Obtain Stable mirth User Shell (user-1 shell)} & $\leftarrow$ & CVE-2023-43208 Exploit Script \\
6 & Internal Port Scanning \& Service Discovery & $\leftarrow$ & mirth User Session \\
7 & Extract Database Connection Info & $\leftarrow$ & mirth.properties \\
8 & Login to Database \& Extract Key Business Data & $\leftarrow$ & Cleartext DB Credentials \\
9 & Offline Cracking of User Credentials & $\leftarrow$ & Username \& Hash from PERSON / PERSON\_PASSWORD Tables \\
10 & {Obtain sedric User Shell via SSH (user-flag shell)} & $\leftarrow$ & Valid Username + Cracked Cleartext Password \\
11 & Analyze Business Forwarding Logic & $\leftarrow$ & XML Config in CHANNEL Table / Local Service Info \\
12 & Identify Controllable Input Points \& Unsafe Fields & $\leftarrow$ & /addPatient Endpoint + Controllable XML Message Body \\
13 & Craft Python Code Execution Payload & $\leftarrow$ & Controllable XML Content + Python Execution Context \\
14 & {Obtain Root via Internal notif Service (root-flag shell)} & $\leftarrow$ & Root-Privileged notif.py Service \\\bottomrule
\end{tabular}
\end{table*}

In contrast, \SysName demonstrates a superior balance between exploratory depth and execution precision. Architecturally, the Executor Agent generates and executes scripts before the Reviewer Agent intervenes to correct course. Consequently, while the statistical ratio of invalid commands may appear comparable to baselines, leveraging the Reviewer to swiftly prune flawed paths minimizes the absolute volume of total commands. This prevents the system from spiraling into resource-draining execution loops while simultaneously achieving the highest shell acquisition rate. A prime example is the Principal machine: \SysName paired with Sonnet 4.5 acquired a root shell utilizing only 79 total commands, with a mere 7 invalid executions, showcasing exceptional payload accuracy. Finally, comparing the foundational models within our framework reveals that Sonnet consistently requires fewer total commands and generates fewer invalid actions than DeepSeek, highlighting how a superior coding model, when properly orchestrated, drastically streamlines the penetration lifecycle.

\section{Classification of Pentesting Failure Reasons}
\label{app:fail}

To systematically analyze the limitations of the evaluated frameworks, we categorize the observed execution failures into the following five distinct taxonomy classes:

\vspace{-\topsep}
\begin{list}{\labelitemi}{\leftmargin=1.5em}
    \setlength{\topsep}{0pt}
    \setlength{\itemsep}{0em}
    \setlength{\parskip}{0pt}
    \setlength{\parsep}{0pt}
    \setlength{\partopsep}{0pt}

    \item \textbf{Planning Error:} Despite possessing sufficient clues or environmental feedback, the agent pursues erroneous or hallucinated attack vectors based on misidentified target assets or flawed vulnerability paths. This causes the overall penetration workflow to deviate significantly from the ground truth without subsequent recovery.

    \item \textbf{Clue Forgetting (Contextual Amnesia):} The agent fails to execute the correct penetration sequence due to the loss or overwriting of previously acquired critical intelligence within its context window (e.g., forgetting user credentials, passwords, open ports, or vulnerability disclosures).

    \item \textbf{Premature Abandonment:} The agent correctly identifies a viable attack trajectory but prematurely halts execution or switches contexts before exploiting the vulnerability.

    \item \textbf{Code/Payload Failure:} Downloaded exploit scripts or autonomously generated payloads fail to execute successfully or achieve the intended objective (e.g., failing to establish Remote Code Execution) due to syntax errors, misconfigurations, or environmental incompatibilities.

    \item \textbf{Information Retrieval Failure:} The agent is unable to acquire the necessary, correct information via web searches or external reconnaissance tools (e.g., failing to locate the ground-truth CVE using SearchSploit).
\end{list}

\end{document}